\documentclass[pra,aps,showpacs,groupedaddress,superscriptaddress,twocolumn,longbibliography,notitlepage]{revtex4-1}
\usepackage[colorlinks=true,linkcolor=blue,urlcolor=blue,citecolor=blue]{hyperref}

\usepackage[utf8x]{inputenc}
\usepackage{color}
\usepackage{bbm} 

\usepackage{amsfonts,amsmath,amssymb,stmaryrd}

\usepackage{graphicx}
\usepackage{subfigure}  
\usepackage{bbm} 
\usepackage{hyperref}
\usepackage[pdftex]{epsfig}
\usepackage{mathrsfs}
\usepackage{verbatim}
\usepackage{centernot}
\usepackage{ulem}
\usepackage{nicefrac} 

\renewcommand{\l}{\left(}
\renewcommand{\r}{\right)}
\newcommand{\hc}{\text{h.c.}}
\newcommand{\eff}{\text{eff}}

\newcommand{\bra}[1]{\langle#1|}
\newcommand{\ket}[1]{|#1\rangle}
\renewcommand{\H}{\hat{\mathcal{H}}}

\renewcommand{\a}{\hat{a}}

\newcommand{\ad}{\hat{a}^\dagger}

\newcommand{\n}{\hat{n}}
\newcommand{\fd}{\hat{f}^\dagger}
\newcommand{\f}{\hat{f}}

\newcommand{\Zt}{\mathbb{Z}_2}

\renewcommand{\ij}{\langle \vec{i}, \vec{j} \rangle}

\usepackage{array}

\usepackage{cancel,ifthen} 
\newcommand{\cmnt}[2][NoInPuT]{\ifthenelse{\equal{#1}{NoInPuT}}{}{{\color{red}\sout{#1}}} {\color{blue} #2}}

\usepackage{bm}	
\renewcommand{\vec}[1]{\bm{#1}}

\begin{document}
\normalem	

\title{Coupling ultracold matter to dynamical gauge fields in optical lattices:\\ From flux-attachment to $\Zt$ lattice gauge theories}

\author{Luca Barbiero}
\affiliation{Center for Nonlinear Phenomena and Complex Systems,
Universit\'e Libre de Bruxelles, CP 231, Campus Plaine, B-1050 Brussels, Belgium}

\author{Christian Schweizer}
\affiliation{Fakult\"at f\"ur Physik, Ludwig-Maximilians-Universit\"at, Schellingstrasse 4, 80799 M\"unchen, Germany}
\affiliation{Max-Planck-Institut f\"ur Quantenoptik, Hans-Kopfermann-Str. 1, 85748 Garching, Germany}

\author{Monika Aidelsburger}
\affiliation{Fakult\"at f\"ur Physik, Ludwig-Maximilians-Universit\"at, Schellingstrasse 4, 80799 M\"unchen, Germany}
\affiliation{Max-Planck-Institut f\"ur Quantenoptik, Hans-Kopfermann-Str. 1, 85748 Garching, Germany}

\author{~\\ Eugene Demler}
\affiliation{Department of Physics, Harvard University, Cambridge, Massachusetts 02138, USA}

\author{Nathan Goldman}
\affiliation{Center for Nonlinear Phenomena and Complex Systems,
Universit\'e Libre de Bruxelles, CP 231, Campus Plaine, B-1050 Brussels, Belgium}

\author{Fabian Grusdt}
\email[Corresponding author email: ]{fabian.grusdt@tum.de}
\affiliation{Department of Physics, Harvard University, Cambridge, Massachusetts 02138, USA}
\affiliation{Department of Physics, Technical University of Munich, 85748 Garching, Germany}

\date{\today}

\maketitle

\textbf{Artificial magnetic fields and spin-orbit couplings have been recently generated in ultracold gases in view of realizing topological states of matter and frustrated magnetism in a highly-controllable environment. Despite being dynamically tunable, such artificial gauge fields are genuinely classical and exhibit no back-action from the neutral particles. Here we go beyond this paradigm, and demonstrate how quantized dynamical gauge fields can be created in mixtures of ultracold atoms in optical lattices. Specifically, we propose a protocol by which atoms of one species carry a magnetic flux felt by another species, hence realizing an instance of flux-attachment. This is obtained by combining coherent lattice modulation techniques with strong Hubbard interactions. We demonstrate how this setting can be arranged so as to implement lattice models displaying a local $\Zt$ gauge symmetry, both in one and two dimensions. We also provide a detailed analysis of a ladder toy model, which features a global $\Zt$ symmetry, and reveal the phase transitions that occur both in the matter and gauge sectors. Mastering flux-attachment in optical lattices envisages a new route towards the realization of strongly-correlated systems with properties dictated by an interplay of dynamical matter and gauge fields.
}

~ \\
\textbf{Introduction}\\
The realization of artificial gauge fields in ultracold gases has further promoted these quantum-engineered systems as versatile quantum simulators~\cite{dalibard2011colloquium,goldman2014light}. While a synthetic magnetic field can be simply introduced by rotating atomic clouds~\cite{cooper2008rapidly}, more sophisticated schemes were developed to generate a wide family of gauge field structures, including spin-orbit couplings~\cite{galitski2013spin} or staggered-flux patterns~\cite{aidelsburger2011experimental,struck2013engineering,jotzu2014experimental}. In fact, the design of magnetic fluxes in optical lattices, through laser-induced tunneling or shaking methods, has been recently exploited in view of realizing topological states of matter~\cite{goldman2016topological,Cooper2018} and frustrated magnetism~\cite{struck2013engineering}. Importantly, these artificial gauge fields are treated as classical and non-dynamical, in the sense that they remain insensitive to the spatial configuration and motion of the atomic cloud:~these engineered systems do not aim to reproduce a complete gauge theory, where particles and gauge fields influence each other. 

In parallel, various theoretical works have suggested several methods by which synthetic gauge fields can be made intrinsically dynamical. A first approach builds on the rich interplay between laser-induced tunneling and strong on-site interactions, which can both be present and finely controlled in an optical lattice~\cite{goldman2014light}:~Under specific conditions, the tunneling matrix elements, which describe the hopping on the lattice but also capture the presence of a gauge field, can become density-dependent~\cite{keilmann2011statistically,greschner2014density,greschner2015anyon,bermudez2015interaction,Strater2016}; see Ref.~\cite{Clark2018} for an experimental implementation of such density-dependent gauge fields. A second approach aims at implementing genuine lattice gauge theories (LGTs), such as the Kogut-Susskind or quantum link models, by directly engineering specific model Hamiltonians through elaborate laser-coupling schemes involving different atomic species and well-designed constraints; see Refs.~\cite{wiese2013ultracold,zohar2015quantum,Dalmonte2016lattice} for reviews and Ref.~\cite{martinez2016real} for an ion-trap realization of the Kogut-Susskind Hamiltonian. Such quantum simulations of LGTs aim to deepen our understanding of fundamental concepts of gauge theories, such as confinement and its interplay with dynamical charges, which are central in high-energy \cite{Fradkin1979} and condensed-matter physics \cite{Lammert1995,Senthil2000,Kitaev2003} and go beyond a mere density-dependence of synthetic gauge fields.

\begin{figure}[t!]
\centering
\epsfig{file=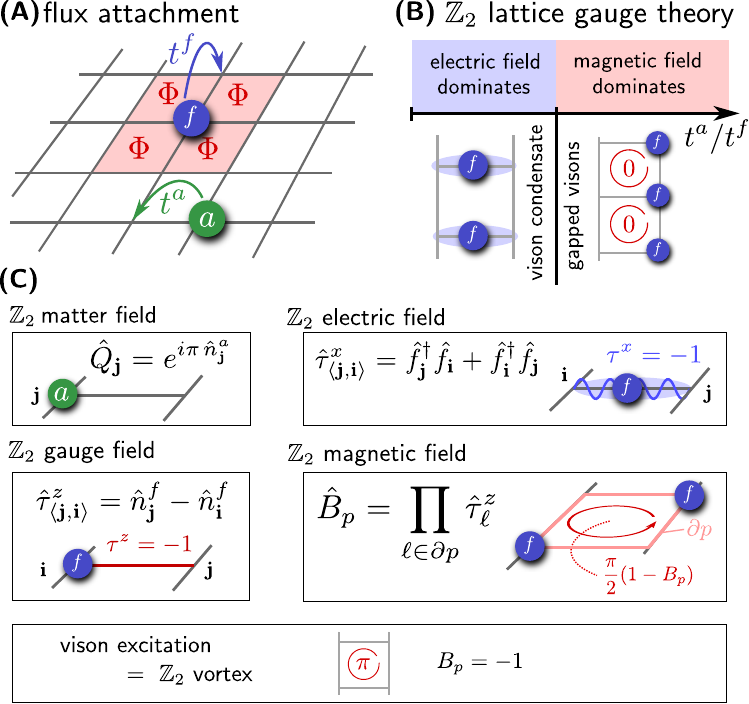, width=0.5\textwidth}
\caption{\textbf{Flux-attachment and dynamical gauge fields with ultracold atoms.} (A) We propose a setup where one atomic species $f$ becomes a source of magnetic flux $\Phi$ (red) for a second species $a$. Both types of atoms undergo coherent quantum dynamics, described by nearest-neighbor tunneling matrix elements $t^{a}$ and $t^f$, respectively. (B) When realized in a ladder geometry, the flux-attachment setup has a $\Zt$ lattice gauge structure. By tuning the ratio of the tunneling elements $t^a / t^f$, we find that the system undergoes a phase transition. The two regimes can be understood in terms of the elementary ingredients of a $\Zt$ LGT, summarized in (C). The matter field $\a$ has a $\Zt$ charge given by the parity of its occupation numbers $\hat{n}^a$. It couples to the $\Zt$ gauge field $\hat{\tau}^z_{\ij}$, defined as the number imbalance of the $f$-particles between different ends of a link. When $|t^a| \ll |t^f|$ the ground state is dominated by tunneling of the $f$-particles, realizing eigenstates of the $\Zt$ electric field $\hat{\tau}^x_{\ij}$. In the opposite limit, $|t^a| \gg |t^f|$, the tunneling dynamics of the $a$-particles prevails and the system realizes eigenstates of the $\Zt$ magnetic field $\hat{B}_p$, defined as a product of the gauge field $\hat{\tau}^z_\ell$ over all links $\ell \in \partial p$ along the edge of a plaquette $p$. The $\Zt$ magnetic field introduces Aharonov-Bohm phases for the matter field, which are $0$ ($\pi$) when the $f$ particles occupy the same (different) leg of the ladder, i.e. if $B_p=1$ ($B_p=-1$). The quantized excitations of the dynamical gauge field correspond to $\Zt$ vortices of the Ising gauge field, so-called visons.}
\label{figSetup}
\end{figure}

In this work, we connect both approaches and demonstrate how LGTs can be realized in ultracold gases through the use of density-dependent gauge fields. As a central ingredient, we devise a scheme to engineer \emph{flux-attachment} for cold atoms moving in an optical lattice. Originally introduced by Wilczek~\cite{wilczek1982magnetic,wilczek1982quantum}, and then widely exploited in the context of the fractional quantum Hall (FQH) effect~\cite{ezawa2008quantum}, flux-attachment is a mathematical construction according to which a certain amount of magnetic-flux quanta is attached to a particle (e.g.~an electron). The resulting composite ``flux-tube-particle" generically satisfies anyonic statistics~\cite{wilczek1982quantum} and naturally appears in field-theoretical formulations of FQH states~\cite{ezawa2008quantum}. Specifically, we show that an optical lattice loaded with two atomic species ($a$ and $f$) can be configured in a way that one species ($f$) becomes a source of magnetic flux $\Phi$ for the other species ($a$):~the magnetic flux is thus effectively attached to moving particles, see Fig.~\ref{figSetup} (A). 

For specific choices of parameters and carefully designed lattice geometries, we demonstrate that this appealing setting can be used to implement interacting quantum systems with local symmetries, realizing $\Zt$ LGTs \cite{Fradkin1979}. These types of models, where the matter field couples to a $\Zt$ lattice gauge field, are especially relevant in the context of high-temperature superconductivity \cite{Lee2008a,Senthil2000} and, more generally, strongly correlated electrons \cite{Sachdev1991,Podolsky2005}. A central question in this context concerns the possibility of a confinement-deconfinement transition in the LGT \cite{Kogut1979}, which would indicate electron fractionalization \cite{Senthil2000,Sedgewick2002,Demler2002}. The proposed model will allow to explore the interplay of a global $U(1)$ symmetry with local $\Zt$ symmetries, which has attracted particular attention in the context of cuprate compounds \cite{Kaul2007,Sachdev2016}. 

Moreover, we will discuss in detail the physics of a toy model characterized by a global $U(1) \times \Zt$ symmetry, which consists of a two-leg ladder geometry and can be directly accessed with state-of-the-art cold-atom experiments. We demonstrate that the toy model features an intricate interplay of matter and gauge fields, as a result of which the system undergoes a phase transition in the $\Zt$ sector depending on the ratio of the species-dependent tunnel couplings $t^a / t^f$, see Fig.~\ref{figSetup} (B). While this transition can be characterized by the spontaneously broken global $\Zt$ symmetry, we argue that an interpretation in terms of the constituents of a $\Zt$ LGT, see Fig.~\ref{figSetup} (C), is nevertheless useful to understand its microscopic origin. We also predict a phase transition of the matter field from an insulating Mott state to a gapless superfluid (SF) regime, associated with the spontaneously broken global $U(1)$ symmetry. For appropriate model parameters, an interplay of both types of transitions can be observed, which resembles the rich physics of higher-dimensional $\Zt$ LGTs at strong couplings. 

The paper is organized as follows. We start by introducing the flux-attachment scheme which is at the heart of the proposed experimental implementation of dynamical gauge fields. Particular attention is devoted to the case of a double-well system, which forms the common building block for realizing $\Zt$ LGTs coupled to matter. Next we study the phase diagram of a toy model with a two-leg ladder geometry, consisting of a matter field coupled to a $\Zt$ gauge field on the rungs. Realistic implementations of the considered models are proposed afterwards, along with a scheme for realizing genuine $\Zt$ LGTs with local instead of global symmetries in two dimensions. This paves the way for future investigations of strongly correlated systems, as discussed in the summary and outlook section. 

~ \\
\textbf{Flux-attachment}\\
The recent experimental implementations of classical gauge fields for ultracold atoms \cite{Aidelsburger2013,Miyake2013,Aidelsburger2014,Kennedy2015,Tai2017} combine two key ingredients \cite{Jaksch2003}: First, the bare tunnel couplings $t$ are suppressed by large energy offsets $|\Delta| \gg t$, realized by a magnetic field gradient or a superlattice potential. Second, tunneling is restored with complex phases $\phi$, by proper time-modulation of the optical lattice \cite{Kolovsky2011a,Goldman2015} at the resonance frequency $\omega = \Delta$ (with $\hbar = 1$ throughout). 

Flux-attachment operates in a strongly-correlated regime, where the energy offsets $\Delta=\omega$ from an external potential are supplemented by inter-species Hubbard interactions of the same magnitude, $U = \omega$ \cite{Chen2011Shaking}. This provides coherent control over the synthetic gauge fields induced by the lattice modulation at frequency $\omega$, see also Refs.~\cite{keilmann2011statistically,greschner2014density,greschner2015anyon,bermudez2015interaction,Strater2016}.

We consider a situation where atoms of a first species, with annihilation operators $\a$, represent a matter field. The atoms of the second type, associated with annihilation operators $\f$, will become the sources of synthetic magnetic flux for the matter field, see Fig.~\ref{figSetup} (A). Namely, the magnetic flux felt by the $a$-particle, as captured by its assisted hopping over the lattice, is only effective in the presence of an $f$-particle. To avoid that -- vice-versa -- the $f$-particles become subject to magnetic flux created by the $a$-particles, static potential gradients affecting only the $f$-particles are used. In the following, we assume that both atomic species are hard-core bosons, although generalizations are possible, for instance when one or both of them are replaced by fermions. 

\begin{figure}[t!]
\centering
\epsfig{file=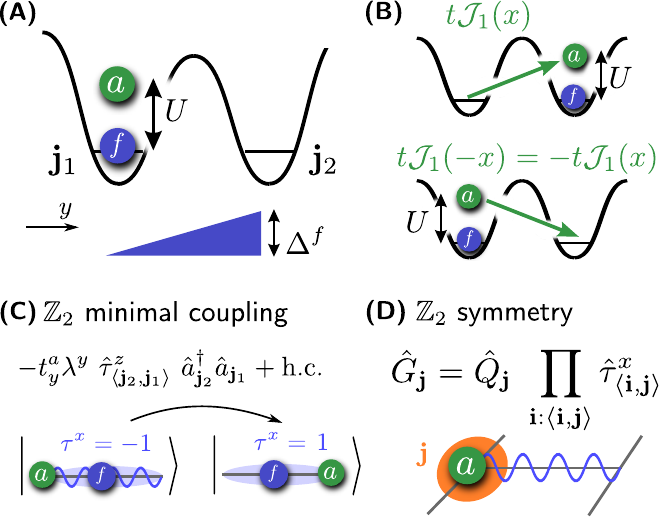, width=0.42\textwidth}
\caption{\textbf{$\Zt$ LGT in a two-well system.} (A) We consider a double-well setup with one atom of each type, $a$ and $f$. Coherent tunneling between the two orbitals at $\vec{j}_1$ and $\vec{j}_2 = \vec{j}_1 + \vec{e}_y$ is suppressed for both species by strong Hubbard interactions $U=\omega$, and for $f$-particles by the energy offset $\Delta^f = \omega$. (B) Tunnel couplings can be restored by resonant lattice modulations with frequency $\omega$. The sign of the restored tunneling matrix element is different when the $a$-particle gains (left panel) or looses (right panel) energy. (C) This difference in sign gives rise to a $\Zt$ gauge structure and allows to implement $\Zt$ minimal coupling of the matter field $\a$ to the link variable defined by the $f$-particles. This term is the common building block for realizing larger systems with a $\Zt$ gauge structure. (D) Such systems are characterized by a symmetry $\hat{G}_{\vec{j}}$ associated with each lattice site $\vec{j}$. Here $\hat{G}_{\vec{j}}$ commutes with the Hamiltonian and consists of the product of the $\Zt$ charge, $\hat{Q}_{\vec{j}} = (-1)^{\hat{n}^a_{\vec{j}}}$, and all electric field lines -- for which $\tau^x=-1$ -- emanating from a volume around site $\vec{j}$ (orange).}
\label{figTwoWell}
\end{figure}

\textbf{Model.} 
The largest energy scale in our problem is set by strong inter-species Hubbard interactions,
\begin{equation}
\H_{\rm int} = U \sum_{\vec{j}}  \n^a_{\vec{j}} \n^f_{\vec{j}},
\label{eqHintDef}
\end{equation}
where $\hat{n}^{a,f}_{\vec{j}}$ denote the density operators of $a$ and $f$-particles on lattice site $\vec{j}$. In order to break the symmetry between $a$- and $f$-particles, we introduce state-dependent static potentials $V_{\alpha}(\vec{j})$, where $\alpha=a,f$. We assume that the corresponding energy offsets between nearest-neighbor (NN) lattice sites $\vec{i}$ and $\vec{j}$ are integer multiples $m^\alpha_{\ij} \in \mathbb{Z}$ of the large energy scale $U$, up to small corrections $|\delta V^\alpha_{\ij}| \ll U$ which are acceptable; namely, 
\begin{equation}
\Delta^\alpha_{\ij} \equiv V_\alpha(\vec{i}) - V_\alpha(\vec{j}) \approx m^\alpha_{\ij} U.
\end{equation}
A minimal example is illustrated in Fig.~\ref{figTwoWell} (A).

Coherent dynamics of both fields are introduced by NN tunneling matrix elements in the $\mu=x,y$ directions, $t^{\alpha}_{\mu}$ respectively. Thus the free part of the Hamiltonian is 
\begin{multline}
\H_0 = - \sum_{\mu=x,y}  \sum_{\ij_\mu} \left[ t_\mu^a \ad_{\vec{j}} \a_{\vec{i}} + t^f_\mu \fd_{\vec{j}} \f_{\vec{i}} + \hc \right] \\
+ \sum_{\vec{j}} \left[ V_a(\vec{j}) ~ \hat{n}^a_{\vec{j}} + V_f(\vec{j}) ~  \hat{n}^f_{\vec{j}} \right]
\label{eqH0def}
\end{multline}
where $\ij_\mu$ denotes a pair of NN sites along direction $\mu$. Tunnel couplings are initially suppressed by the external potentials $\Delta^\alpha = m^\alpha U$ and the strong Hubbard interactions,
\begin{equation}
U \gg |t^\alpha_{x,y}|.
\end{equation}

To restore tunnel couplings with complex phases we include a time-dependent lattice modulation,
\begin{equation}
\H_\omega(t) = \sum_{\vec{j}}  ~ V_\omega(\vec{j},t) \l \ad_{\vec{j}} \a_{\vec{j}} + \fd_{\vec{j}} \f_{\vec{j}} \r.
\label{eqHomegaDef}
\end{equation}
It acts equally on both species and is periodic in time, $V_\omega(\vec{j},t + 2 \pi / \omega) = V_\omega(\vec{j},t)$, with frequency $\omega = U$ resonant with the inter-species interactions. Summarizing, our Hamiltonian is
\begin{equation}
\H(t) = \H_0 + \H_{\rm int} + \H_\omega(t).
\label{eqHtComplete}
\end{equation}

\textbf{Effective hopping Hamiltonian.} 
From now on we consider resonant driving, $U = \omega \gg |t_{\mu}^\alpha|$, where the lattice modulation $\H_\omega(t)$ in Eq.~\eqref{eqHomegaDef} restores, or renormalizes, all tunnel couplings of $a$- and $f$-particles. As derived in the Supplements (SM), we obtain an effective hopping Hamiltonian to lowest order in $1/\omega$,
\begin{multline}
\H_{\rm eff} = - \sum_{\mu=x,y}  \sum_{\ij_\mu} \biggl[ t^a_\mu ~ \ad_{\vec{i}} \a_{\vec{j}} ~ \hat{\lambda}^\mu_{\ij_\mu}~ e^{i \hat{\varphi}^\mu_{\ij_\mu}} + \hc \\
 + t^f_\mu ~ \fd_{\vec{i}} \f_{\vec{j}} ~ \hat{\Lambda}^\mu_{\ij_\mu}~ e^{i \hat{\theta}^\mu_{\ij_\mu}} + \hc \biggr].
\label{eqHeffGeneral}
\end{multline}
The Hermitian operators $\hat{\lambda}^\mu_{\ij_\mu}$ and $\hat{\varphi}^\mu_{\ij_\mu}$ (respectively $\hat{\Lambda}^\mu_{\ij_\mu}$ and $\hat{\theta}^\mu_{\ij_\mu}$) in Eq.~\eqref{eqHeffGeneral} describe the renormalization of the tunneling amplitudes and phases, for $a$ (resp. $f$) particles; they are mutually commuting and depend only on the number imbalance $\hat{n}^f_{\vec{j}}-\hat{n}^f_{\vec{i}}$ (resp. $\hat{n}^a_{\vec{j}} - \hat{n}^a_{\vec{i}}$) associated with the respective complementary species. Our result in Eq.~\eqref{eqHeffGeneral} is reminiscent of the models discussed in Ref.~\cite{bermudez2015interaction}.

Explicit expressions for $\hat{\lambda}$, $\hat{\varphi}$, $\hat{\Lambda}$, $\hat{\theta}$ can be obtained by considering their matrix elements on the relevant many-body states $\ket{\psi_r}$ and $\ket{\psi_s}$ in the Fock basis that are involved in the various hopping processes. For an $a$-particle transitioning from state $\ket{\psi_s}$ to $\ket{\psi_r}$, corresponding to a relative potential and / or interaction energy offset $\Delta_{r s} = n_{r s} ~ \omega$ with integer $n_{r s} \in \mathbb{Z}$, the matrix elements are given by
\begin{equation}
\bra{\psi_r} \ad_{\vec{i}} \a_{\vec{j}} ~ \hat{\lambda}^\mu_{\ij_\mu} \ket{\psi_s} = |\mathcal{J}_{n_{r s}}(x)|.
\label{eqDeflbda}
\end{equation}
Here $\mathcal{J}_n$ denotes the Bessel function of the first kind, $x=A_{\vec{i},\vec{j}} / \omega$ is the dimensionless driving strength, and
\begin{equation}
V_\omega(\vec{i},t) - V_\omega(\vec{j},t) = A_{\vec{i},\vec{j}} \cos \l \omega t + \phi_{\vec{i}, \vec{j}}\r.
\label{eqDefVomega}
\end{equation}
Without loss of generality, we assume $\omega, A_{\vec{i},\vec{j}} > 0$ throughout the paper. 

The complex phases of the restored tunnelings are also determined by the many-body energy offsets $\Delta_{r s} = n_{r s} \omega$. If $n_{r s} \geq 0$ the particle gains energy in the hopping process and 
\begin{equation}
\bra{\psi_r} \ad_{\vec{i}} \a_{\vec{j}} ~ \hat{\varphi}^\mu_{\ij_\mu} \ket{\psi_s} = |n_{r s}| \phi_{\vec{i}, \vec{j}}.
\label{eqDefPhiPos}
\end{equation}
In contrast, if $n_{r s}<0$ the particle looses energy and 
\begin{equation}
\bra{\psi_r} \ad_{\vec{i}} \a_{\vec{j}} ~ \hat{\varphi}^\mu_{\ij_\mu} \ket{\psi_s} = |n_{r s}| ( \pi - \phi_{\vec{i}, \vec{j}} ).
\label{eqDefPhiNeg}
\end{equation}
In this case there is an additional $n_{r s} \pi$ phase shift due to the reflection properties of the Bessel function, $J_n(-x) = (-1)^n J_n(x)$, see Fig.~\ref{figTwoWell} (B). This $n_{r s} ~ \pi$ phase shift is at the core of the LGT implementations discussed below. Similar results are obtained for $\hat{\Lambda}^\mu_{\ij_\mu}$ and $\hat{\theta}^\mu_{\ij_\mu}$ by exchanging the roles of $a$ and $f$, see SM. Note, however, that the symmetry between $a$ and $f$ can be broken by a careful design of the potentials $V_a$ and $V_f$, and this will be exploited in the next paragraph.

As illustrated in Fig.~\ref{figSetup} (A), our scheme allows to implement effective Hamiltonians [Eq.~\eqref{eqHeffGeneral}] describing a mixture of two species, where one acts as a source of magnetic flux for the other, see also Ref.~\cite{bermudez2015interaction}. A detailed discussion of the resulting Harper-Hofstadter model with dynamical gauge flux is provided in the SM. By analogy with the physics of the FQH effect \cite{Prange1990,Jain1990}, we expect that this flux-attachment gives rise to interesting correlations, and possibly to quasiparticle excitations with non-trivial statistics.

\textbf{$\Zt$ LGT in a double-well.} 
Now we apply the result in Eq.~\eqref{eqHeffGeneral} and discuss a minimal setting, where one $a$ and one $f$-particle tunnel between the two sites $\vec{j}_1$ and $\vec{j}_2 = \vec{j}_1 + \vec{e}_y$ of a double-well potential, see Fig.~\ref{figTwoWell} (A); $\vec{e}_y$ denotes the unit vector along $y$. This system forms the central building block for the implementation of $\Zt$ LGTs in larger systems, proposed below. We assume $V_a(\vec{j}_{i}) \equiv 0$ for $i=1,2$ but introduce a potential offset $V_f(\vec{j}_2) = \Delta^f + V_f(\vec{j}_1)$ for the $f$ species, breaking the symmetry between $a$- and $f$-particles.

\emph{Effective Hamiltonian.--}
For $\Delta^f = U = \omega$ and lattice modulations with a trivial phase $\phi_{\vec{j}_1,\vec{j}_2} = 0$, the effective Floquet Hamiltonian in Eq.~\eqref{eqHeffGeneral} becomes
\begin{equation}
\H_{\rm eff}^{\rm 2well} = - t^a_y ~ \lambda^y~ \hat{\tau}^z_{\langle \vec{j}_2,\vec{j}_1 \rangle} \l   \ad_{\vec{j}_2} \a_{\vec{j}_1}  + \hc \r - t_y^f ~ \hat{\Lambda} ~ \hat{\tau}^x_{\langle \vec{j}_2,\vec{j}_1 \rangle},
\label{eqDefHeff2}
\end{equation}
with notations defined as follows. We describe the $f$-particle by a pseudo spin $\nicefrac{1}{2}$,
\begin{equation}
\hat{\tau}^z_{\langle \vec{j}_2,\vec{j}_1 \rangle} = \hat{n}^f_{\vec{j}_2} - \hat{n}^f_{\vec{j}_1}, \qquad \hat{n}^f_{\vec{j}_2} + \hat{n}^f_{\vec{j}_1} = 1,
\label{eqDefLinkVar}
\end{equation}
which becomes a link variable in the $\Zt$ LGT, see Fig.~\ref{figSetup} (C). The Pauli matrix $\hat{\tau}^x_{\langle \vec{j}_2,\vec{j}_1 \rangle} = ( \fd_{\vec{j}_2} \f_{\vec{j}_1} + \hc )$ describes tunneling of the $f$-particle. 

As shown in Fig.~\ref{figTwoWell} (B), the interaction energy of the matter field changes by $\pm U$ in every tunneling event. As a result the amplitude renormalization in Eq.~\eqref{eqDefHeff2} is $\lambda^y = |\mathcal{J}_1(\nicefrac{A_{\vec{j}_2,\vec{j}_1}}{\omega})|$, see Eq.~\eqref{eqDeflbda}, and the phase of the restored tunnel couplings is $e^{i \hat{\varphi}} = \hat{\tau}^z_{\langle \vec{j}_2,\vec{j}_1 \rangle}$ by Eqs.~\eqref{eqDefPhiPos}, \eqref{eqDefPhiNeg}. Because the $f$-particle is subject to an additional potential offset $\Delta^f = U$ between the two sites, its energy can only change by $0$ or $2 U$ in a tunneling event. Hence the phase of the restored tunneling in Eq.~\eqref{eqDefHeff2} is trivial, $\hat{\theta} = 0$ as in Eqs.~\eqref{eqDefPhiPos}, \eqref{eqDefPhiNeg}, but the amplitude renormalization 
\begin{equation}
\hat{\Lambda} = \mathcal{J}_0 \l \nicefrac{A_{\vec{j}_2,\vec{j}_1}}{\omega} \r \hat{n}^a_{\vec{j}_1} + \mathcal{J}_2 \l \nicefrac{A_{\vec{j}_2,\vec{j}_1}}{\omega} \r \hat{n}^a_{\vec{j}_2}, 
\label{eqHeff2s2pLbdaMain}
\end{equation}
depends on the configuration of the $a$-particle in general. 

The effective Hamiltonian \eqref{eqDefHeff2} realizes a minimal version of a $\Zt$ LGT: the link variable $\hat{\tau}^z_{\langle \vec{j}_2,\vec{j}_1 \rangle} \simeq e^{i \pi \hat{\mathcal{A}}}$ provides a representation of the dynamical $\Zt$ gauge field $\hat{\mathcal{A}}$, which is quantized to $0$ and $1$. The corresponding $\Zt$ electric field is given by the Pauli matrix $\hat{\tau}^x_{\langle \vec{j}_2,\vec{j}_1 \rangle}$, defining electric field lines on the link. The $\Zt$ charges $\hat{Q}_{\vec{j}_i}$, defined on the two sites $\vec{j}_i$ with $i=1,2$, are carried by the $a$-particle, $\hat{Q}_{\vec{j}_i} = \exp ( i \pi \hat{n}^a_{\vec{j}_i} )$. These ingredients are summarized in Fig.~\ref{figSetup} (C) and justify our earlier notion that the $a$ and $f$-particles describe matter and gauge-fields, respectively. The Hamiltonian in Eq.~\eqref{eqDefHeff2} realizes a minimal coupling \cite{Kogut1979} of the $a$-particles to the gauge field, see Fig.~\ref{figTwoWell} (C).

\emph{Symmetries.--}
Each of the two lattice sites $\vec{j}_i$ is associated with a $\Zt$ symmetry. The operators generating the $\Zt$ gauge group in the double-well system,
\begin{equation}
\hat{g}_{i} = \hat{Q}_{\vec{j}_i} ~ \hat{\tau}^x_{\langle \vec{j}_2,\vec{j}_1 \rangle}, \qquad i=1,2,
\label{eqDefZtgaugeGroup2wll}
\end{equation}
both commute with the effective Hamiltonian in Eq.~\eqref{eqDefHeff2},  $[\hat{g}_{i} , \H_{\rm eff}^{\rm 2well} ] = 0$ for $i=1,2$. This statement is not entirely trivial for the first term in Eq.~\eqref{eqDefHeff2}: While $\hat{\tau}^z_{\langle \vec{j}_2,\vec{j}_1 \rangle}$ and $ \ad_{\vec{j}_2} \a_{\vec{j}_1}$ do not commute with $\hat{\tau}^x_{\langle \vec{j}_2,\vec{j}_1 \rangle}$ and $\hat{Q}_{\vec{j}_{i}}$ individually, their product commutes with $\hat{g}_{i}$. The second term in Eq.~\eqref{eqDefHeff2} trivially commutes with $\hat{g}_i$ because $[\hat{\Lambda}, \hat{Q}_{\vec{j}_i}] = 0$, see Eq.~\eqref{eqHeff2s2pLbdaMain}. 

Physically, Eq. \eqref{eqDefZtgaugeGroup2wll} establishes a relation between the $\Zt$ electric field lines, $\tau^x_{\langle \vec{j}_2,\vec{j}_1 \rangle} = -1$, and the $\Zt$ charges from which they emanate, see Fig.~\ref{figTwoWell} (D). Note that the eigenvalues of $\hat{g}_1$ and $\hat{g}_2$ are not independent, because $\hat{g}_1 \hat{g}_2=-1$ for the considered case with a single $a$ particle tunneling in the double-well system. 

The model in Eq.~\eqref{eqDefHeff2} is invariant under the gauge symmetries $\hat{g}_{i}$ for all values of the modulation strength $A_{\vec{j}_2,\vec{j}_1}$. In general, both terms in the effective Hamiltonian couple the $\Zt$ charge to the gauge field. An exception is obtained for lattice modulation strengths $\nicefrac{A_{\vec{j}_2,\vec{j}_1}}{\omega} = x_{02} $ for which
\begin{equation}
\mathcal{J}_0(x_{02}) = \mathcal{J}_2(x_{02}).
\label{eqJ02cond}
\end{equation}
In this case, neither of the amplitude renormalizations
\begin{flalign}
\hat{\Lambda} \to \Lambda_{02} &= \mathcal{J}_0(x_{02}) \approx 0.32, \\
 \lambda^y =  \lambda_{02} &= \mathcal{J}_1(x_{02}) \approx 0.58,
\end{flalign} 
is operator valued, and the second term in the Hamiltonian only involves the $\Zt$ gauge field. The weakest driving for which Eq.~\eqref{eqJ02cond} is satisfied has $x_{02} \approx 1.84$.

\begin{figure*}[t!]
\centering
\epsfig{file=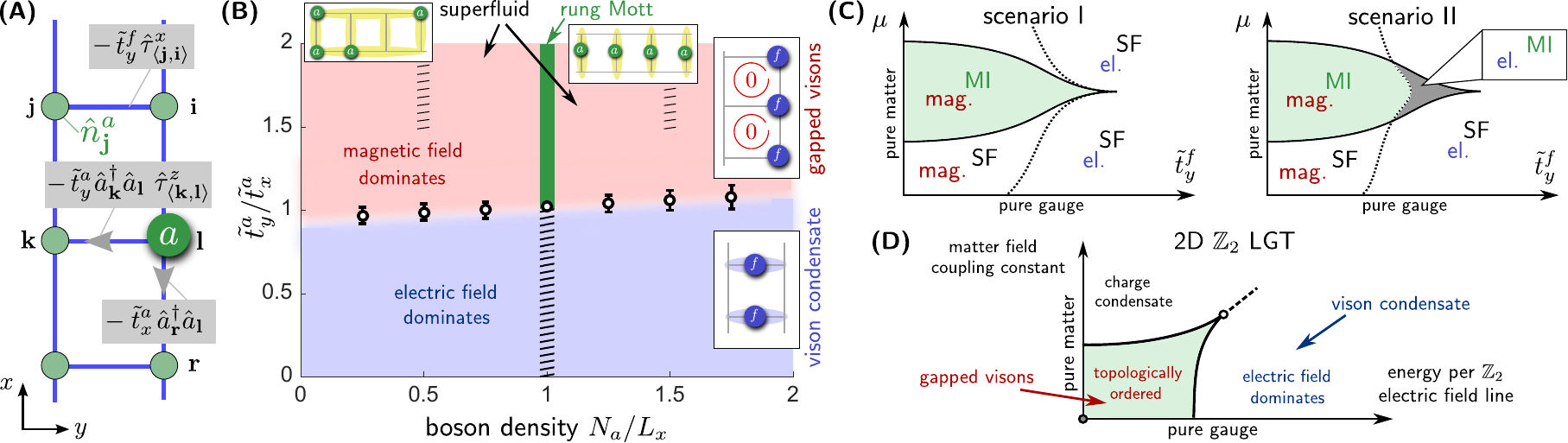, width=0.99\textwidth}
\caption{\textbf{Coupling matter to a $\Zt$ gauge field in a two-leg ladder.} (A) We consider the Hamiltonian \eqref{eqDefHLGT} describing $a$-particles which are minimally coupled to the $\Zt$ gauge field $\hat{\tau}^z_{\ij_y}$ on the rungs of a two-leg ladder. (B) The phase diagram, obtained by DMRG simulations at $\tilde{t}^f_y / \tilde{t}^a_x = 0.54$, contains a SF-to-Mott transition in the charge sector at a commensurate density of the matter field, $N_a = L_x$. In addition, we find a transition in the gauge sector, from an ordered region with a broken global $\Zt$ symmetry where the $\Zt$ magnetic field dominates and the vison excitations of the gauge field are gapped (red), to a disordered regime where the $\Zt$ electric field is dominant and visons are strongly fluctuating in a condensed state (blue). This behavior is reminiscent of confinement-deconfinement transitions in higher-dimensional LGTs. Along the hatched lines at commensurate fillings $N_a/L_x=1/2,1,3/2$, insulating charge-density wave states could exist, but conclusive numerical results are difficult to obtain. (C) The conjectured schematic phase diagram of Eq.~\eqref{eqDefHLGT} is shown in the $\mu-\tilde{t}^f_y$ plane, where $\mu$ denotes the chemical potential for $\a$ particles and $2 \tilde{t}^f_y$ corresponds to the energy cost per $\Zt$ electric field line along a rung. Our numerical data is consistent with two scenarios: In I, the interplay of gauge and matter fields prevents a fully disordered Mott phase, whereas the latter exists in scenario II. The behavior in scenario I resembles the phase diagram of the more general 2D $\Zt$ LGT \cite{Fradkin1979,Lammert1995,Senthil2000,Kitaev2003} sketched in (D). In our DMRG simulations here, as well as in the following figures, we keep up to $1400$ DMRG states with $5$ finite-size sweeps; the relative error on the energies is kept smaller than $10^{-7}$.}
\label{figZ2LGT}
\end{figure*}

~ \\
\textbf{Matter-gauge field coupling in two-leg ladders}\\
In the following we study the physics of coupled matter and gauge fields in a two-leg ladder, accessible with numerical density-matrix-renormalization-group (DMRG) simulations \cite{WhiteDMRG}. Our starting point is a model with minimal couplings to the $\Zt$ gauge field on the rungs of the ladder, which is characterized by a global $U(1) \times \Zt$ symmetry, see Fig.~\ref{figZ2LGT} (A). Here we study its phase diagram. As explained later, the model can be implemented in existing ultracold atom setups by simply coupling multiple double-well systems. Generalizations to extended ladder models with local symmetries and minimal couplings to a $\Zt$ gauge field on all links are discussed in the SM. 

\textbf{The model.}
We combine multiple double-well systems \eqref{eqDefHeff2} to a two-leg ladder, by introducing tunnelings $t_x^a$ of the matter field along $x$. Further, we impose that the $f$-particles can only move along the rungs, $t_x^f = 0$, and each rung is occupied by one $f$-particle. Thus, we can continue describing the $f$ degrees of freedom by link variables $\hat{\vec{\tau}}_{\ij_y}$ as defined in Eq.~\eqref{eqDefLinkVar}. The number of $a$-particles $N_a$ will be freely tunable.

\emph{Effective Hamiltonian.--}
For a properly designed configuration of lattice gradients and modulations, presented in detail later, we obtain an effective Hamiltonian
\begin{multline}
\H_{\rm 2leg} = - \sum_{\ij_x} \l  t^a_x ~ \hat{\lambda}^x_{\ij_x}  \ad_{\vec{j}} \a_{\vec{i}}  + \hc \r  \\
-  \sum_{\ij_y} \bigg[ t^a_y  ~\lambda^y \l \ad_{\vec{j}} \a_{\vec{i}} \hat{\tau}^z_{\ij_y} + \hc \r + t_y^f ~  \hat{\Lambda}^y_{\ij_y} \hat{\tau}^x_{\ij_y} \bigg].
\label{eqHeffLadder}
\end{multline}
Expressions for the amplitude renormalizations $\lambda^y \in \mathbb{R}$ and $\hat{\lambda}^x$, $\hat{\Lambda}^y$ are provided in the SM. 

For the specific set of driving strengths $x=x_{02}$ that we encountered already in the double well problem, see Eq.~\eqref{eqJ02cond}, we find that $\hat{\Lambda}^y$ only has a weak dependence on the $\Zt$ charges, $\hat{Q}_{\vec{j}} = (-1)^{\hat{n}^a_{\vec{j}}}$. Similarly, the amplitude renormalization $\hat{\lambda}^x$ depends weakly on the $\Zt$ magnetic field $\hat{B}_p$ only; here
\begin{equation}
\hat{B}_p = \prod_{\ij_y \in \partial p} \hat{\tau}^z_{\ij_y}
\label{eqDefBp}
\end{equation}
is defined as a product over all links $\ij_y$ on the rungs belonging to the edge $\partial p$ of plaquette $p$. Hence, for these specific modulation strengths,
\begin{flalign}
[\hat{\Lambda}^y_{\ij_y} , \hat{Q}_{\vec{l}}] &= [\hat{\Lambda}^y_{\ij_y} , \hat{\vec{\tau}}_{\langle \vec{k}, \vec{l} \rangle}] = 0, \label{eqLbdaYcommQx02} \\
[\hat{\lambda}^x_{\ij_x} , \hat{B}_{p}] &= [\hat{\lambda}^x_{\ij_x} , \a^{(\dagger)}_{\vec{l}}] = 0.  \label{eqLbdaYcommBx02}
\end{flalign}

\emph{Symmetries.--}
Now we discuss the symmetries of the effective Hamiltonian \eqref{eqHeffLadder} at the specific value of the driving strengths $x_{02}$. In the case of decoupled rungs, i.e. for $t^a_x = 0$, every double-well commutes with $\hat{g}_i$, $i=1,2$ from Eq.~\eqref{eqDefZtgaugeGroup2wll}. These symmetries are no longer conserved for $t^a_x \neq 0$; in this general case a global $\Zt$ symmetry remains: 
\begin{equation}
\hat{V}_i = \prod_{j=1}^{L_x} \hat{g}_{i}(j \vec{e}_x), \qquad i=1,2,
\label{eqZtSymm}
\end{equation}
with $\hat{g}_{i}(j \vec{e}_x) = (-1)^{\hat{Q}_{j \vec{e}_x + (i-1) \vec{e}_y}} \hat{\tau}^x_{ \langle j \vec{e}_x + \vec{e}_y, j \vec{e}_x \rangle_y }$ and for which $\hat{V}_i^2 =1$. Using Eqs.~\eqref{eqLbdaYcommQx02}, \eqref{eqLbdaYcommBx02} one readily confirms that  $[\H_{\rm 2leg}, \hat{V}_i] = 0$ for $i=1,2$.

Summarizing, the effective model is characterized by the global $U(1)$ symmetry associated with the conservation of the number of $a$ particles, and the global $\Zt$ symmetry $\hat{V}_1$. Note that the second $\Zt$ symmetry, $\hat{V}_2$, follows as a consequence of combining $\hat{V}_1$ with the global $U(1)$ symmetry: By performing the global $U(1)$ gauge transformation $\a_{\vec{j}} \to - \a_{\vec{j}}$ for all sites $\vec{j}$, $\hat{V}_2$ is obtained from $\hat{V}_1$. Thus, the overall symmetry is $U(1) \times \Zt$. 

\emph{Physical constituents.--}
In the following, we will describe the physics of the ladder models using the ingredients of $\Zt$ LGTs, see Fig.~\ref{figSetup} (C). The quantized excitations of the $\Zt$ lattice gauge field are vortices of the $\Zt$ (or Ising) lattice gauge field, so-called visons \cite{Senthil2000}. They are defined on the plaquettes of the ladder: If the plaquette term in Eq.~\eqref{eqDefBp} is $B_p=1$, there is no vison on $p$; the presence of an additional $\Zt$ flux, $B_p=-1$, corresponds to a vison excitation on plaquette $p$. Since the matter field $\a$ couples to the $\Zt$ gauge field, the resulting interactions with the visons determine the phase diagram of the many-body Hamiltonian, as in higher-dimensional $\Zt$ LGTs \cite{Senthil2000,Kitaev2003}. 

\textbf{Quantum phase transitions of matter and gauge fields.}
We start from the microscopic model in Eq.~\eqref{eqHeffLadder} and simplify it by making a mean-field approximation for the renormalized tunneling amplitudes, which depend only weakly on $\hat{Q}_{\vec{j}}$ and $\hat{B}_p$. Replacing them by $\mathbb{C}$-numbers, $\tilde{t}^a_{x} = t^a_x \langle \hat{\lambda}^x \rangle$, $\tilde{t}^f_y = t^f_y \langle \hat{\Lambda}^y \rangle$ and $\tilde{t}^a_y = t^a_y \lambda^y$ leads to the conceptually simpler Hamiltonian,
\begin{multline}
\H_{\rm 2leg}^{\rm simp} = -  \sum_{\ij_x}  \tilde{t}^a_x \l \ad_{\vec{j}} \a_{\vec{i}}  + \hc \r \\
- \sum_{\ij_y} \left[ \tilde{t}^a_y \l \ad_{\vec{j}} \a_{\vec{i}}  \hat{\tau}_{\ij_y}^z + \hc \r  + \tilde{t}^f_y ~ \hat{\tau}_{\ij_y}^x \right],
\label{eqDefHLGT}
\end{multline}
illustrated in Fig.~\ref{figZ2LGT} (A). Later, by introducing a more sophisticated driving scheme, we will show that this model can be directly implemented using ultracold atoms. The simpler Hamiltonian \eqref{eqDefHLGT} has identical symmetry properties as Eq.~\eqref{eqHeffLadder}. Now we analyze Eq.~\eqref{eqDefHLGT} by means of the DMRG technique. In the phase diagram we find at least three distinct phases, resulting from transitions in the gauge- and matter-field sectors, see Fig.~\ref{figZ2LGT} (B). Here we describe their main features; for more details the reader is referred to the SM.

\begin{figure*}[t!]
\centering
\epsfig{file=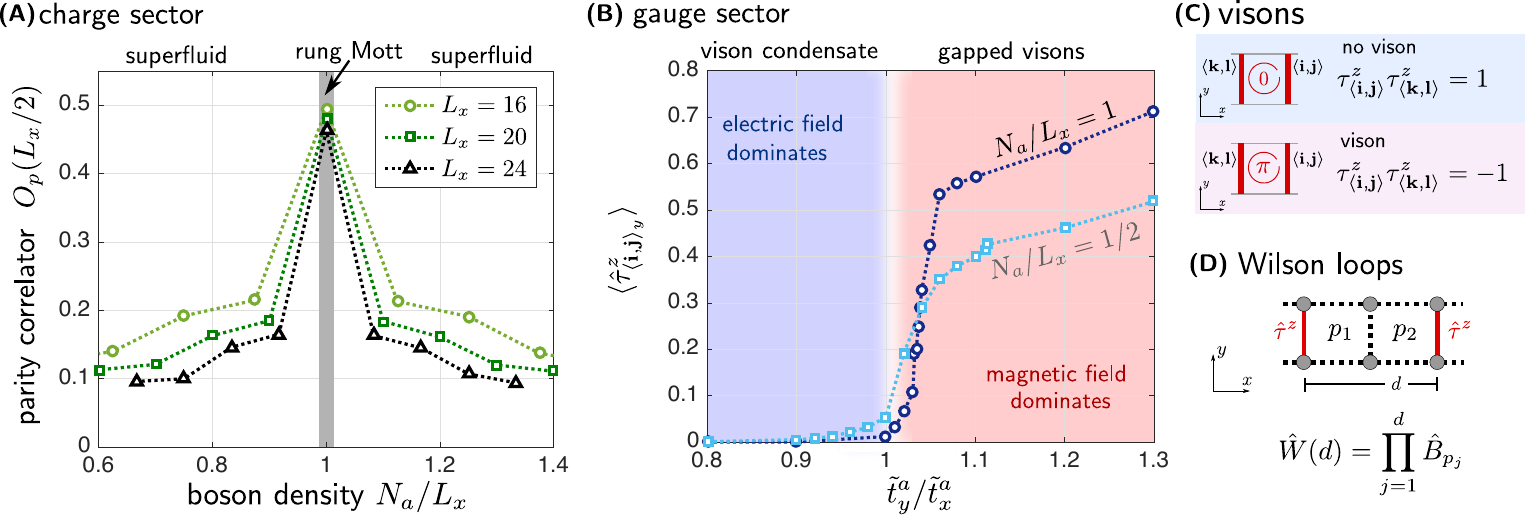, width=0.99\textwidth}
\caption{\textbf{Characterizing phase transitions of matter coupled to a $\Zt$ gauge field in a two-leg ladder.} (A) In the charge sector we observe transitions from a SF state, characterized by a vanishing parity correlator $O_p(L_x \to \infty) \to 0$ in the thermodynamic limit, to an insulating rung-Mott state at the commensurate filling $N_a = L_x$, characterized by $O_p(L_x \to \infty) > 0$ and exponentially decaying correlations. Here we present exemplary results for $\tilde{t}^a_y/\tilde{t}^a_x = 3$ and $\tilde{t}^f_y / \tilde{t}^a_x = 0.54$. (B) In the gauge sector we find a transition from a disordered phase, where the $\Zt$ electric field dominates, to a phase where the $\Zt$ magnetic field dominates. In the second case, the order parameter $\langle \hat{\tau}_{\ij_y}^z \rangle \neq 0$ corresponds to a spontaneously broken global $\Zt$ symmetry \eqref{eqZtSymm}. In the two phases, the corresponding vison excitations of the $\Zt$ gauge field (C) have different characteristics. The numerical results in (A) [respectively (B)] are obtained by considering periodic boundary conditions [respectively $L_x=96$ rungs with open boundaries]. (D) Analogues of Wilson loops $\hat{W}(d)$ in the two-leg ladder are string operators of visons.}
\label{figPhaseTrans}
\end{figure*}

\emph{Transition in the matter sector.--}
First we concentrate on the conceptually simpler phase transition taking place in the charge sector. When the tunneling along the legs is weak, $\tilde{t}^a_x \lesssim \left[ (\tilde{t}^f_y)^2 + (\tilde{t}^a_y)^2 \right]^{1/2} - \tilde{t}^f_y$, and the number $N_a$ of $a$-particles is tuned, we observe a pronounced transition from a SF to a rung-Mott phase \cite{Crepin2011} at the commensurate filling $N_a = L_x$, where $L_x$ denotes the total number of rungs in the system. Similarly to the analysis in Refs.~\cite{Berg2008,Endres2011,Fazzini2017}, this transition can be captured by the parity operator 
\begin{equation}
 O_p(l)= \left\langle \exp \left[ i \pi \sum_{j<l} \l \hat{n}^a_{j \vec{e}_x} +  \hat{n}^a_{j \vec{e}_x + \vec{e}_y} - \frac{N_a}{L_x} \r \right]  \right\rangle.
 \label{eqDefOP}
\end{equation}
In the limit $l \simeq L_x$ and $L_x \to \infty$ this observable $O_p$ remains finite only in the Mott insulating regime. Our results in Fig.~\ref{figPhaseTrans} (A) confirm that $O_p$ takes large values with a weak size dependence for $N_a = L_x$. On the other hand when $N_a/L_x \neq 1$ is slightly increased or decreased, the parity $O_p$ suddenly becomes smaller and a significant $L_x$ dependence is observed which is consistent with a vanishing value in thermodynamic limit.

For larger values of $\tilde{t}^a_x$ (see SM for details), no clear signatures of a Mott phase are found: The parity operator $O_p$ takes significantly smaller values, the calculated Mott gap becomes a small fraction of $\tilde{t}^a_x$, consistent with a finite-size gap, and we checked that the decay of two-point correlations follows a power-law at long distances until edge-effects begin to play a role. Because of the global $U(1)$ symmetry of the model, a possible SF-to-Mott transition in the quasi-1D ladder geometry would be of Berezinskii-Kosterlitz-Thouless (BKT) type. Hence the gap would be strongly suppressed and the correlation length exponentially large, making it impossible to determine conclusively from our numerical results if the ground state is a gapped Mott state or not. Indeed, for single-component hard-core bosons on a two-leg ladder it has been shown by bosonization that an infinitesimal inter-leg coupling is sufficient to open up an exponentially small Mott gap \cite{Crepin2011,Piraud2015}.

Similar considerations apply at the commensurate fillings $N_a / L_x = 1/2, 3/2$, where previous work on single-component bosons in a two-leg ladder \cite{Piraud2015} pointed out the possible emergence of an insulating charge density wave (CDW) for large enough $\tilde{t}^a_y/\tilde{t}^a_x$ [hatched areas in Fig.~\ref{figZ2LGT} (B)]. Our numerical results indicate that such CDWs may exist in this regime also in our model, Eq.~\eqref{eqDefHLGT}, but a more accurate analysis is required in order to properly locate the transition point.

\emph{Transition in the gauge sector.--}
Next we focus on the gauge sector, described by the $f$-particles, in which we observe a phase transition when tuning the ratios of the tunnel couplings. For example, in Fig.~\ref{figPhaseTrans} (B) we tune $\tilde{t}^a_y/\tilde{t}^a_x$ while keeping $\tilde{t}^f_y / \tilde{t}^a_x$ fixed. We find a transition from a regime where $\langle \hat{\tau}_{\ij_y}^z \rangle = 0$ to a region with a non-vanishing order parameter $\langle \hat{\tau}_{\ij_y}^z \rangle \neq 0$. Similar behavior is obtained when tuning $\tilde{t}_y^f/\tilde{t}_y^a$ while keeping $\tilde{t}^a_x/\tilde{t}_y^a$ fixed, see SM for more details. 

The observed transition is associated with a spontaneous breaking of the global $\Zt$ symmetry \eqref{eqZtSymm} of the model. The $f$-particles go from a regime where they are equally distributed between the legs, $\langle \hat{\tau}_{\ij_y}^z \rangle = 0$, to a two-fold degenerate state with population imbalance, $\langle \hat{\tau}_{\ij_y}^z \rangle \neq 0$. Such behavior occurs in the insulating and SF regimes of the charge sector, and it is only weakly affected by the filling value $N_a/L_x$, see Fig.~\ref{figZ2LGT} (B). Our numerical results in Fig.~\ref{figPhaseTrans} (B) indicate that the transition is continuous. The critical exponent is in good agreement with a value of $1/8$ as expected for an Ising universality class, especially in the insulating regime, but it is also possible that the transition is of BKT type associated with the opening of a gap in the gauge sector.

The two phases of the $\Zt$ gauge field are easily understood in the limiting cases. When $\tilde{t}^a_y = 0$, the ground state is an eigenstate of the $\Zt$ electric field $\hat{\tau}^x_{\ij_y}$ on the rungs, with eigenvalues $1$. The $\Zt$ magnetic field is strongly fluctuating and there exist no $\Zt$ electric flux loops. Thus also the vison number is strongly fluctuating, and the state can be understood as a vison condensate. In the opposite limit, when $\tilde{t}^a_y \to \infty$, the kinetic energy of the matter field dominates. In this case the $\Zt$ magnetic field is effectively static, and its configuration is chosen in order to minimize the kinetic energy of the $a$-particles. This is achieved when the effective Aharonov-Bohm phases on the plaquettes vanish, i.e. for $\hat{B}_p=1$, see Fig.~\ref{figSetup} (C). In this case vison excitations with $\hat{B}_p=-1$, see Fig.~\ref{figPhaseTrans} (C), correspond to localized defects in the system, which cost a finite energy corresponding to the vison gap. 

Lattice gauge theories with local instead of global symmetries are characterized by Wilson loops \cite{Kogut1979}. Their closest analogues in our two-leg ladder model are string operators of visons, 
\begin{equation}
W(d) = \prod_{j=1}^d \langle \hat{B}_{p_j} \rangle = \langle \hat{\tau}^z_{\ij_y} \hat{\tau}^z_{\langle \vec{i}+d \vec{e}_x , \vec{j} +d \vec{e}_x \rangle_y} \rangle
\label{eqDefWilson}
\end{equation}
see Fig.~\ref{figPhaseTrans} (D). In the disordered phase (electric field dominates) we found numerically that $W(d) \to 0$ when $d \to \infty$, whereas $W(d)$ remains finite at large distances in the ordered phase (magnetic field dominates), see SM. 

This qualitatively different behavior of the Wilson loop is reminiscent of the phenomenology known from the ubiquitous confinement-deconfinement transitions found in $(2+1)$ dimensional LGTs \cite{Kogut1979,Fradkin1979,Senthil2000}: There, visons are gapped in the deconfined phase and the Wilson loop decays only weakly exponentially with a perimeter law; in the confining phase, visons condense and the Wilson loop decays much faster with an exponential area law. Although the ordered phase which we identified in the two-leg ladder geometry is characterized by a spontaneously broken global $\Zt$ symmetry, this analogy suggests that it represents a precursor of the genuine deconfined phase expected in 2D $\Zt$ LGTs with local symmetries. 

\emph{Interplay of matter and gauge fields.--}
Finally, we discuss the interplay of the observed phase transitions in the gauge and matter sectors. To this end we find it convenient to consider the phase diagram in the $\mu-\tilde{t}^f_y$ plane, where $\mu$ denotes a chemical potential for the $a$-particles and $\tilde{t}^f_y$ controls fluctuations of the $\Zt$ electric field. We collect our result in the schematic plots in Fig.~\ref{figZ2LGT} (C): Deep in the SF phase, realized for small $\mu$ and $N_a/L_x < 1$, $\tilde{t}^f_y$ drives the transition in the gauge sector. Because of a particle-hole symmetry of the hard-core bosons in the model, similar results apply for large $\mu$ and $N_a/L_x > 1$. On the other hand, when $\tilde{t}^f_y$ is small, permitting a sizable Mott gap at commensurate fillings, $\mu$ drives the SF-to-Mott transition. 

More interesting physics can happen at the tip of the Mott lobe, for commensurate fillings $N_a = L_x$. This corresponds to the hatched regime in Fig.~\ref{figZ2LGT} (B), where we cannot say conclusively, if the system is in a gapped Mott phase. To obtain better understanding of the commensurate regime, we first argue that a SF cannot co-exist with the ordered phase of the gauge field at commensurate fillings: In this regime the $\Zt$ gauge field acquires a finite expectation value, $\langle \tau^z_{\ij_y} \rangle \neq 0$. This leads to a term in the Hamiltonian $\sim - \tilde{t}^a_y \langle \tau^z_{\ij_y} \rangle \ad_{\vec{i}} \a_{\vec{j}}$, which is expected to open a finite Mott gap, following the arguments in Refs.~\cite{Crepin2011,Piraud2015}. Therefore, only the two scenarios shown in Fig.~\ref{figZ2LGT} (C) are possible: In the first case, the Mott insulator co-exists only with the ordered phase of the gauge field; in the second scenario the Mott state co-exists with the disordered phase of the gauge field. 

To shed more light on this problem, we consider the case when the Mott gap $\Delta$ is much larger than the tunneling $\tilde{t}^a_x$. When $\tilde{t}^a_x=0$, every rung represents an effective localized spin-$1/2$ degree of freedom. As shown in the SM, finite tunnelings $\tilde{t}^a_x \ll \Delta$ introduce anti-ferromagnetic couplings between these localized moments, and in this limit our system can be mapped to an XXZ chain. It has an Ising anisotropy and the ground state has a spontaneously broken $\Zt$ symmetry everywhere, except when $\tilde{t}^f_y / \tilde{t}^a_y \to \infty$ where an isotropic Heisenberg model is obtained and the ground state has power-law correlations. The transition from the gapped Mott state, corresponding to the ordered phase of the $\Zt$ gauge field, to a symmetric state of two decoupled SFs with a disordered gauge field, is of BKT type~\cite{Giamarchi2003}. 

Although our last argument is limited to small values of $\tilde{t}^a_x$, it indicates that scenario I in Fig.~\ref{figZ2LGT} (C) may be more likely, but more detailed investigations will be required to draw a final conclusion. At least in the limit of small $\tilde{t}^a_x$ and large couplings $\tilde{t}^f_y$ of the gauge field, our analysis proofs that there exists an intricate interplay of the phase transitions in the gauge and matter sectors. Such behavior, characteristic for scenario I in Fig.~\ref{figZ2LGT} (C), is reminiscent of the phase diagram of the 2D $\Zt$ LGT \cite{Fradkin1979,Kitaev2003}, see Fig.~\ref{figZ2LGT} (D). In that case, the phase at weak couplings has topological order as in Kitaev's toric code \cite{Kitaev2003}, and the disordered phases are continuously connected to each other at strong couplings.

~ \\
\textbf{Implementations: coupled double-well systems}\\
Now we describe how the models discussed above, and extensions thereof, can be implemented in state-of-art ultracold atom setups. The double-well system introduced around Eq.~\eqref{eqDefHeff2} constitutes the building block for implementing larger systems with a $\Zt$ gauge symmetry, or even genuine $\Zt$ LGTs, because it realizes a minimal coupling of the matter field to the gauge field \cite{Kogut1979}, see Fig.~\ref{figTwoWell} (C). We start by discussing the two-leg ladder Hamiltonian $\H_{\rm 2leg}$, Eq.~\eqref{eqHeffLadder}; then we present a scheme, based on flux-attachment, for implementing a genuine $\Zt$ LGT coupled to matter in a 2D square lattice.

\begin{figure}[b!]
\centering
\epsfig{file=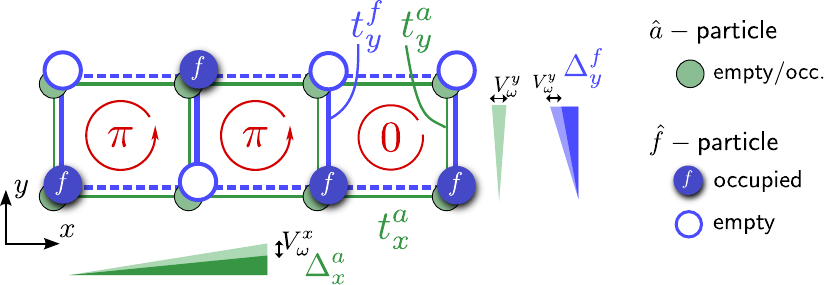, width=0.48\textwidth}
\caption{\textbf{Implementing matter-gauge field coupling in a two-leg ladder.} 
Multiple double-well systems as described in Fig.~\ref{figTwoWell} are combined to form a two-leg ladder by including hopping elements $t^a_x$ of the $a$-particles along the $x$ direction. Coherent tunneling is first suppressed by strong inter-species Hubbard interactions $U$ and static potential gradients: $\Delta_x^a = U$ for $a$-particles along $x$, and $\Delta^f_y=U$ for $f$-particles along $y$. The tunnel couplings are restored by a resonant lattice shaking with frequency $\omega = U$, realized by a modulated potential gradient $V_\omega(\vec{j},t) = ( j_x V_\omega^x + j_y V_\omega^y ) \cos(\omega t)$ seen by both species. We assume that each rung is occupied by exactly one $f$-particle, which can thus be described by a link variable, while the number $N_a$ of $a$-particles is freely tunable. As shown in the SM, the special choice for the driving strengths $V_\omega^x / \omega = V_\omega^y / \omega = x_{02}$ leads to an effective Hamiltonian with matter coupled to $\Zt$ lattice gauge fields on the rungs. The gradient $\Delta^a_x = U$ guarantees that the $a$-particles pick up only trivial phases $\hat{\varphi}^x=0$ while tunneling along the legs of the ladder. Hence the Aharonov-Bohm phases (red) associated with the matter field become $0$, or $\pi$ corresponding to a vison excitation. They are determined by the plaquette terms $\hat{B}_p$ defined in Eq.~\eqref{eqDefBp}, reflecting the configuration of $f$-particles.}
\label{figTwoLegLadder}
\end{figure}

\textbf{Two-leg ladder geometry.}
The ladder system shown in Fig.~\ref{figZ2LGT} (A) can be obtained by combining multiple double-wells \eqref{eqDefHeff2} and introducing tunnelings $t_x^a$ of the matter field along $x$, while $t^f_x = 0$. The lattice potential is modulated along $y$ with amplitude $A_{\vec{j_2},\vec{j_1}} = V_\omega^y$, as in the case of a single double-well. As described in Fig.~\ref{figTwoLegLadder}, we introduce an additional static potential gradient with strength $\Delta^a_x = U = \omega$ per lattice site along $x$ and modulate it with frequency $\omega$ and amplitude $V_\omega^x$. 

As shown in the SM, this setup leads to the effective Hamiltonian \eqref{eqHeffLadder}. For the specific set of driving strengths $V_\omega^x / \omega = V_\omega^y / \omega = x_{02}$, see Eq.~\eqref{eqJ02cond}, the amplitude renormalizations are $\lambda^y = \lambda_{02}$ and 
\begin{multline}
\hat{\lambda}^x_{\ij_x} = \frac{1}{2} \l 1 - \hat{\tau}^z_{\langle \vec{i} \pm \vec{e}_y , \vec{i} \rangle}  \hat{\tau}^z_{\langle \vec{j} \pm \vec{e}_y , \vec{j} \rangle}
\r  \mathcal{J}_0(x_{02})\\
+ \frac{1}{2} \l 1 +  \hat{\tau}^z_{\langle \vec{i} \pm \vec{e}_y , \vec{i} \rangle}  \hat{\tau}^z_{\langle \vec{j} \pm \vec{e}_y , \vec{j} \rangle}
\r  \mathcal{J}_1(x_{02}).
\label{eqDeflbdaxMain}
\end{multline}

\emph{Simplified model.--}
Now we discuss a further simplification of the model in Eq.~\eqref{eqHeffLadder}, leaving its symmetry group unchanged. We note that, even for the specific choice of the driving strengths $V_\omega^x / \omega = V_\omega^y / \omega = x_{02}$, the renormalized tunnel couplings of the $a$-particles along $x$ still depend explicitly on the $\Zt$ gauge fields on the adjacent rungs, see Eq.~\eqref{eqDeflbdaxMain}. This complication can be avoided, by simultaneously modulating the gradient along $x$ at two frequencies, $\omega$ and $2 \omega$, with amplitudes $V_\omega^x$ and $V_{2\omega}^x$; i.e. we consider the following driving term in Eq.~\eqref{eqHomegaDef},
\begin{equation}
V_\omega(\vec{j},t) = [ j_x V_\omega^x + j_y V_\omega^y ] \cos(\omega t) + j_x V_{2 \omega}^x  \cos(2 \omega t).
\label{eqMultiFreqDriving}
\end{equation}

Following Ref.~\cite{Goldman2015} we obtain expressions for the restored tunnel couplings along $x$ for an energy offset $n \omega$ introduced by the Hubbard interactions $U = \omega$ between $a$ and $f$-particles; $\lambda_n = \sum_{\ell=- \infty}^{\infty} \mathcal{J}_{n - 2 \ell}(x^{(1)}) \mathcal{J}_{\ell}(x^{(2)} /2 )$ where $x^{(1)} = V_\omega^x / \omega$ and $x^{(2)} = V_{2\omega}^x / \omega$ (see SM). By imposing the conditions $\lambda_0 = \lambda_1 = \lambda_2$, we obtain a simplified effective Hamiltonian where $\hat{\lambda}^x_{\ij_x} \to \lambda^x \in \mathbb{R}$ is no longer operator-valued, and thus completely independent of the $\Zt$ gauge field $\hat{\tau}^z$. The weakest driving strengths for which this condition is met is given by
\begin{equation}
x^{(1)} = x^{(1)}_{012} \approx 1.71, \qquad  x^{(2)} = x^{(2)}_{012} \approx 1.05,
\label{eqx1012x2012}
\end{equation}
where $\lambda^x = \lambda_{012} \approx 0.37$. A similar approach can be used to make $\hat{\Lambda}^y$ independent of the $\Zt$ charges, which allows to implement $\H_{\rm 2leg}^{\rm simp}$ from Eq.~\eqref{eqDefHLGT}.

\textbf{Realizing a $\Zt$ LGT in a 2D square lattice.}
Now we present a coupling scheme of double-wells which results in an effective 2D LGT Hamiltonian with genuine local symmetries, in addition to the global $U(1)$ symmetry associated with $a$-number conservation. We will derive a model with $\Zt$ gauge-invariant minimal coupling terms $\sim \hat{\tau}^z_{\ij} \ad_{\vec{j}} \a_{\vec{i}}$ along all links of the square lattice. 

\begin{figure}[t!]
\centering
\epsfig{file=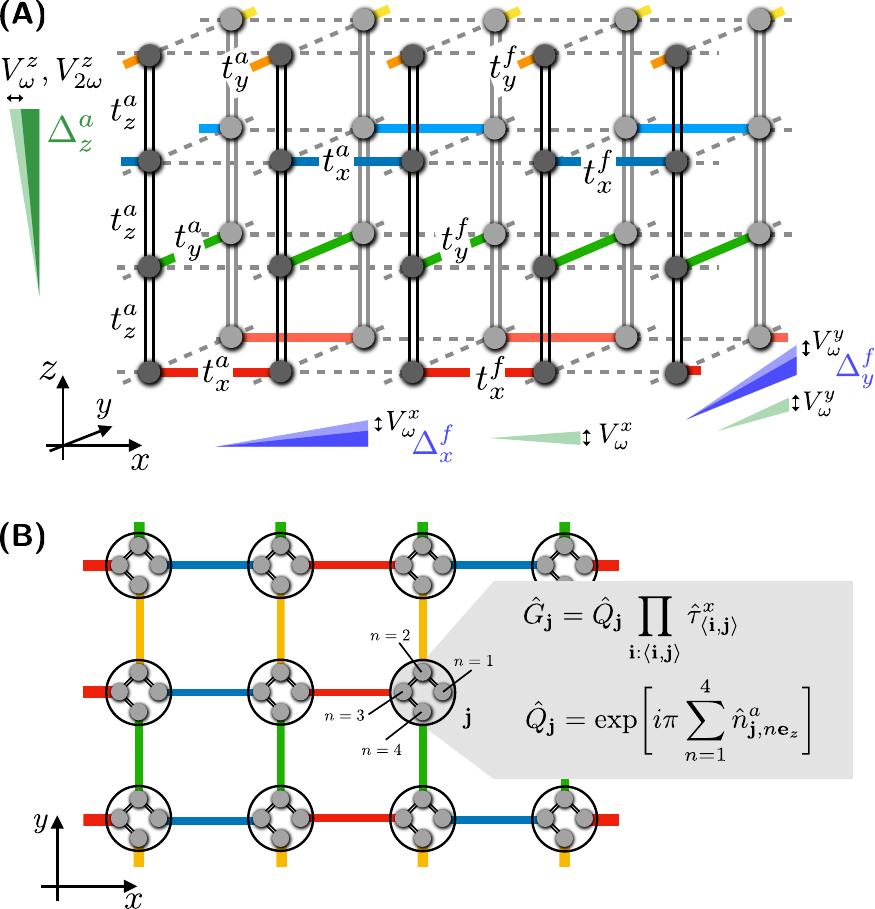, width=0.48\textwidth}
\caption{\textbf{Realizing $\Zt$ LGT coupled to matter in 2D.} (A) Multiple double-well systems as described in Fig.~\ref{figTwoWell} are combined in the shown brick-wall lattice. Each of its four layers along $z$-direction is used to realize one of the four links connecting every lattice site of the 2D square lattice (B) to its four nearest neighbors. The double-well systems are indicated by solid lines (colors), and they are only coupled by tunnelings of $a$-particles along the $z$-direction, with amplitudes $t^a_z$. (B) The restored hopping Hamiltonian $\H_{\rm 2DLGT}$ in the 2D lattice has local symmetries $\hat{G}_{\vec{j}}$ associated with all lattice sites $\vec{j}$, i.e. $[ \H_{\rm 2DLGT} ,\hat{G}_{\vec{j}}] = 0$.}
\label{fig2DZ2LGT}
\end{figure}

\emph{Setup.--}
We consider the setup shown in Fig.~\ref{fig2DZ2LGT} (A) in a layered 2D optical lattice, which is a particular type of brick-wall lattice. The $a$-particles tunnel vertically between the layers in $z$ direction, with coupling matrix element $t^a_z$, and along the links indicated in the figure with tunnel couplings $t^a_x$ and $t^a_y$. Every tube consisting of four lattice sites with coordinates $x$, $y$ and $n \vec{e}_z$ for $n=1,2,3,4$ defines a super-site $\vec{j} = x \vec{e}_x + y \vec{e}_y$ in the effective 2D lattice shown in Fig.~\ref{fig2DZ2LGT} (B). The four links connecting every super-site to its nearest neighbors $\vec{i}: \ij$ are realized by double-well systems, with exactly one $f$-particle each, in different layers of the optical lattice. The $f$-particles are only allowed to tunnel between the sites of their respective double-wells in the $x-y$ plane, with amplitudes $t^f_x$ and $t^f_y$, while tunneling along $z$-direction is suppressed, $t^f_z = 0$.

For the realization of the individual double-well systems, we consider a modulated potential gradient along $x$ and $y$, seen equally by the matter and gauge fields. The modulation amplitudes $V_\omega^x / \omega = V_\omega^y / \omega = x_{02}$ are chosen to simplify the amplitude renormalization of $f$-particle tunneling. As previously, we consider static potential gradients along $x$ and $y$ directions of $\Delta^f_x=\Delta^f_y = U$ per site, seen only by the $f$-particles, and work in a regime where $U = \omega \gg |t^\nu_\mu|$, with $\mu=x,y,z$ and $\nu = a, f$.
 
To realize $a$-particle tunneling along $z$ which is independent of the $\Zt$ gauge fields $\hat{\tau}^z$ on the links in the $x-y$ plane, we add a static potential gradient of $\Delta^a_z$ per site along $z$-direction. It is modulated by two frequency components $\omega$ and $2 \omega$, with amplitudes $V_{\omega}^z$ and $V_{2 \omega}^z$. These driving strengths are chosen as in Eq.~\eqref{eqx1012x2012}, i.e. $V_{\omega}^z / \omega = x^{(1)}_{012}$ and $V_{2 \omega}^z / \omega = x^{(2)}_{012}$, such that the restored tunnel couplings with amplitude $t^a_z \lambda_{012}$ become independent of the $f$-particle configuration.

\emph{Effective Hamiltonian.--}
Combining our results from the previous section, we obtain the effective hopping Hamiltonian $\H_{\rm 2DLGT}$ for the setup described in Fig.~\ref{fig2DZ2LGT},
\begin{multline}
\H_{\rm 2DLGT} = - t^f_{xy} \sum_{\ij}  \hat{\Lambda}_{\ij}  ~ \hat{\tau}^x_{\ij}  \\
-  t^a_{xy} \lambda_{02} \sum_{\ij} \l  \hat{\tau}^z_{\ij} \ad_{\vec{i},m_{\ij}} \a_{\vec{j},m_{\ij}} + \hc \r \\
- t_z^a \lambda_{012} \sum_{\vec{j}} \sum_{n=1}^3 \l \ad_{\vec{j},n+1} \a_{\vec{j},n}  + \hc \r
\label{eqHeff2D}
\end{multline}
using the same notation as introduced earlier. Here we treat the $z$-coordinate $n \vec{e}_z$, with $n=1,...,4$, as an internal degree of freedom, while $\vec{j}$ is a site index in the 2D square lattice; $m_{\ij} \in \{1,2,3,4\}$ denotes the $z$-coordinate corresponding to double-well $\ij$. For simplicity we assumed that $t^a_x = t^a_y = t^a_{xy}$ and $t^f_x = t^f_y = t^f_{xy}$. The amplitude renormalization for $f$-particles in the $x$-$y$ plane depends on the $\Zt$ charges $\hat{Q}_{\vec{j},n}$ (see SM),
\begin{multline}
 \hat{\Lambda}_{\ij} = \frac{1}{2} \left[ \mathcal{J}_0(x_{02}) + \mathcal{J}_1(x_{02}) \right] \\
 + \hat{Q}_{\vec{i},m_{\ij}} \hat{Q}_{\vec{j},m_{\ij}}  \frac{1}{2} \left[ \mathcal{J}_1(x_{02}) - \mathcal{J}_0(x_{02}) \right].
 \label{eqDefLBDA2dZ2LGT}
\end{multline}
Using the multi-frequency driving scheme explained around Eq.~\eqref{eqMultiFreqDriving}, a situation where $\hat{\Lambda}_{\ij}$ becomes independent of the $\Zt$ charges can be realized.

A simplified effective Hamiltonian, where the internal degrees of freedom are eliminated, can be obtained when $U = \omega \gg t^a_z $ and $\lambda_{012} t^a_z \gg t^a_{xy}$; the first inequality is required by the proposed implementation scheme. In this limit, the tunneling of $a$-particles along $z$ can be treated independently of the in-plane tunnelings $t^a_{xy}$. The ground state with a single $a$-particle tunneling along $z$ at super-site $\vec{j}$ is $\ad_{\vec{j}} \ket{0}$, where $\ad_{\vec{j}} =  \sum_{n=1}^4 \phi_n \ad_{\vec{j},n}$ with $\phi_1 = \phi_4 = (5 + \sqrt{5})^{-1/2}$ and $\phi_2 = \phi_3 = (1 + 1/\sqrt{5})^{1/2} / 2$. It is separated by an energy gap $\Delta \varepsilon = \lambda_{012} t^a_z \gg t^a_{xy}$ from the first excited state, which justifies our restriction to this lowest internal state. 

The ground state energy $\varepsilon_{2 a}$ with two hard-core $a$-particles tunneling along $z$ in the same super-site is larger than twice the energy $\varepsilon_{a}$ of a single $a$-particle, by an amount $U_{\eff}$, i.e. $\varepsilon_{2 a} = 2 \varepsilon_{a} + U_{\eff}$. By solving the one and two-particle problems exactly, we find $U_{\eff} = \lambda_{012}  t^a_z$. In the effective model restricted to the lowest internal state, this offset corresponds to a repulsive Hubbard interaction on the super-sites $\vec{j}$. Because $U_{\eff} \gg t^a_{xy}$, double occupancy of super-sites is strongly suppressed, and we can treat the new operators $\a_{\vec{j}}^{(\dagger)}$ as hard-core bosons. 

By projecting the Hamiltonian \eqref{eqHeff2D} to the lowest internal state on every super-site, we arrive at the following simplified model,
\begin{multline}
\H_{\rm 2DLGT}^{\rm simp} = \varepsilon_{a} \sum_{\vec{j}} \ad_{\vec{j}} \a_{\vec{j}} - t^f_{xy} \sum_{\ij}  \hat{\Lambda}_{\ij}  ~ \hat{\tau}^x_{\ij}  \\
-  t^a_{xy} \lambda_{02} |\phi_1|^2 \sum_{\ij \in {\rm E}} \l  \hat{\tau}^z_{\ij} \ad_{\vec{i}} \a_{\vec{j}} + \hc \r \\
-  t^a_{xy} \lambda_{02} |\phi_2|^2 \sum_{\ij \in {\rm B}} \l  \hat{\tau}^z_{\ij} \ad_{\vec{i}} \a_{\vec{j}} + \hc \r.
\label{eqHZ2LGT2D}
\end{multline}
Here we distinguish between two sets of links, $\ij \in {\rm E}$ or ${\rm B}$, which are realized in layers at the edge $n=1,4$ (${\rm E}$) and in the bulk $n=2,3$ (${\rm B}$) in the 3D implementation, see Fig.~\ref{fig2DZ2LGT} (A). Because the internal state has different weights $|\phi_1|^2 \approx 0.14$ and $|\phi_2|^2 \approx 0.36$, they are associated with different tunneling amplitudes. This complication can be avoided by realizing bare tunnelings of $a$-particles with different strengths on ${\rm E}$ and ${\rm B}$-type bonds. 

\emph{Symmetries.--}
In contrast to the two-leg ladder \eqref{eqHeffLadder}, the models in Eqs.~\eqref{eqHeff2D}, \eqref{eqHZ2LGT2D} are both characterized by local $\Zt$ gauge symmetries. The $\Zt$ charge on a super-site is defined as $\hat{Q}_{\vec{j}} = \exp [i \pi \sum_{n=1}^4 \hat{n}^a_{\vec{j},n}]$, which becomes $\hat{Q}_{\vec{j}} = \exp [i \pi \ad_{\vec{j}} \a_{\vec{j}}]$ when projected to the lowest internal state. The $\Zt$ gauge group is generated by
\begin{equation}
\hat{G}_{\vec{j}} = \hat{Q}_{\vec{j}} \prod_{\vec{i}: \langle \vec{j}, \vec{i} \rangle} \hat{\tau}^x_{\langle \vec{j}, \vec{i} \rangle},
\label{eqDefGjLdr2D}
\end{equation}
where the product on the right includes all links $\langle \vec{j}, \vec{i} \rangle$ connected to site $\vec{j}$.  

It holds $[ \H_{\rm 2DLGT} , \hat{G}_{\vec{j}} ]=0$ and $[\H_{\rm 2DLGT}^{\rm simp}, \hat{G}_{\vec{j}}]=0$ for all $\vec{j}$, using the respective $\Zt$ charge operators. These results follow trivially for the first line of Eqs.~\eqref{eqHeff2D}, \eqref{eqHZ2LGT2D} which contain only the operators $\hat{\tau}^x_{\ij}$ and $\hat{n}_{\vec{j},n}^a$ ($\hat{n}_{\vec{j}}$), see also Eq.~\eqref{eqDefLBDA2dZ2LGT}. For the last two lines in the effective Hamiltonians, it is confirmed by a straightforward calculation.

In addition to the local $\Zt$ gauge invariance, the models \eqref{eqHeff2D}, \eqref{eqHZ2LGT2D} have a global $U(1)$ symmetry associated with the conservation of the $a$-particle number. Very similar Hamiltonians have been studied in the context of strongly correlated electrons, where fractionalized phases with topological order have been identified \cite{Sedgewick2002}. When the $a$-particles condense, effective models without the global $U(1)$ symmetry can also be realized. These are in the same symmetry class as Kitaev's toric code \cite{Kitaev2003}.

~ \\
\textbf{Discussion}\\
We have presented a general scheme for realizing flux-attachment in 2D optical lattices, where one species of atoms becomes a source of magnetic flux for a second species. For a specific set of parameters, we demonstrated that the effective Floquet Hamiltonian describing our system has a $\Zt$ gauge structure. This allows to implement experimentally a dynamical $\Zt$ gauge field coupled to matter using ultracold atoms, as we have shown specifically for a double-well setup, two-leg ladders and in a 2D geometry. Because our scheme naturally goes beyond one spatial dimension, the $\Zt$ magnetic field -- and the corresponding vison excitations -- play an important role in our theoretical analysis of the ground state phase diagram. Moreover, the link variables in our system are realized by particle-number imbalances on neighboring sites, making experimental implementations of our setup feasible using existing platforms (as described e.g. in Refs.~\cite{Miyake2013,Aidelsburger2014,Tai2017}).  

Our theoretical analysis of hard-core bosons coupled to a $\Zt$ gauge field in a ladder geometry revealed a SF-to-Mott transition in the charge sector, as well as a transition in the gauge sector. The latter is characterized by a spontaneously broken global $\Zt$ symmetry, but we argued that it can be understood as a precursor of the confinement-deconfinement transitions which are ubiquitous in LGTs, high-energy physics and strongly correlated quantum many-body system. Leveraging the powerful toolbox of quantum gas microscopy, our approach paves the way for new studies of LGTs with full resolution of the quantum mechanical wavefunction. This is particularly useful for analyzing string \cite{Endres2011,Hilker2017} and topological \cite{Grusdt2016TP} order parameters, which are at the heart of LGTs but difficult to access in more conventional settings.

As we have demonstrated, extensions of our LGT setting to 2D systems with local rather than global symmetries are possible. Here we propose a realistic scheme to implement a genuine $\Zt$ LGT with minimal coupling of the matter to the gauge-field on all links of a square lattice. On the one hand, this realizes one of the main ingredients of Kitaev's toric code \cite{Kitaev2003,Paredes2008,Dai2017} -- a specific version of a LGT coupled to matter, which displays local $\Zt$ gauge symmetry and hosts excitations with non-Abelian anyonic statistics. On the other hand, the systems that can be implemented with our technique are reminiscent of models studied in the context of nematic magnets \cite{Lammert1993,Lammert1995,Podolsky2005} and strongly correlated electron systems \cite{Senthil2000,Sedgewick2002,Demler2002}. Other extensions of our work include studies of more general systems with flux-attachment, which are expected to reveal physics related to the formation of composite fermions in the FQH effect. 

In terms of experimental implementations, we restricted our discussion in this article to ultracold atom setups. Other quantum simulation platforms, such as arrays of superconducting qubits \cite{Houck2012}, provide promising alternatives however. Generalizations of our scheme to these systems are straightforward, and a detailed analysis of the feasibility of our proposal in such settings will be devoted to future work.

\newpage

\section*{Supplementary Material}
\noindent
\textbf{I. Implementing dynamical gauge fields}\\
Here we describe in detail how synthetic gauge fields with their own quantum dynamics can be realized, and implemented using ultracold atoms. We begin by quickly reviewing results for the case of a single particle in a double-well potential, which we use later on to derive the effective Hamiltonian in a many-body system.

\textbf{Single-particle two-site problem.} 
We consider the following Hamiltonian describing a single particle hopping between sites $\ket{1}$ and $\ket{2}$,
\begin{equation}
\H_2 = - t \l \ket{2}\bra{1} + \ket{1}\bra{2} \r + \l \Delta_{2,1} + \Delta^\omega_{2,1}(t) \r \ket{2}\bra{2}.
\label{eqHbtoy}
\end{equation}
Here $t>0$ denotes the bare tunnel coupling which is strongly suppressed by the energy offset $|\Delta_{2,1}| \gg t$. Tunneling is then restored by a modulation 
\begin{equation}
\Delta^\omega_{2,1}(t) = A_{2,1} \cos \l \omega t + \phi_{2,1} \r.
\label{eqBondModToy}
\end{equation}

For resonant shaking, $\omega = \Delta_{2,1}$, it has been shown in Ref.~\cite{Kolovsky2011a} that the dynamics of Eq.~\eqref{eqHbtoy} can be described by the following effective Hamiltonian,
\begin{equation}
\H_{2,\rm eff} = - \tilde{t} \l  \ket{2}\bra{1} e^{i \phi_{2,1}} + \ket{1}\bra{2} e^{-i \phi_{2,1}}  \r.
\label{eqHeffToySM}
\end{equation}
The amplitude of the restored tunneling is given by
\begin{equation}
\tilde{t} = t ~ \mathcal{J}_1 \l \nicefrac{A_{2,1} }{ \omega } \r
\end{equation}
and the complex phase $\phi_{2,1}$ is determined directly from the modulating potential $\Delta^\omega_{2,1}(t)$.

\begin{figure}[t!]
\centering
\epsfig{file=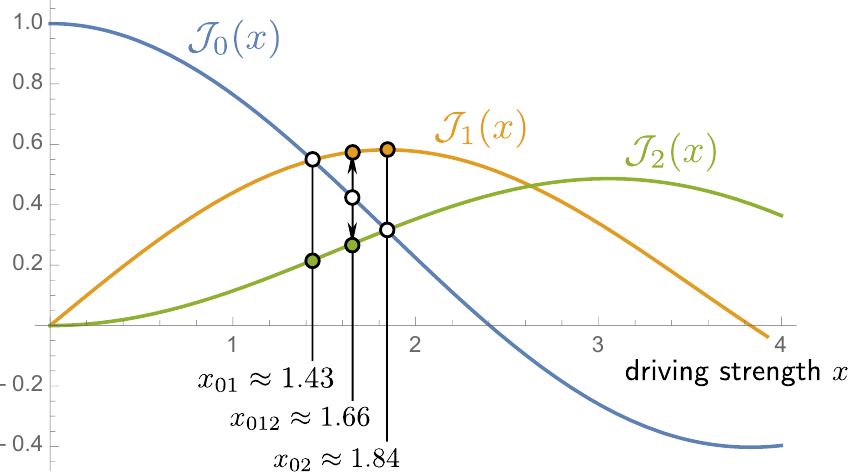, width=0.43\textwidth}
\caption{\textbf{Renormalized tunneling amplitudes determined by Bessel functions.} The tunnelings of $a$ and $f$-particles, which are initially suppressed by energy offsets $n \omega$ for integer $n$, are restored by resonant lattice modulations. Their renormalized amplitude is proportional to $\mathcal{J}_n(x)$, where $x$ is the dimensionless driving strength $A/\omega$. For $x \approx x_{012} = 1.66$ the first three Bessel functions are closest to one another, $\mathcal{J}_1(x) \approx \mathcal{J}_0(x) \approx \mathcal{J}_2(x)$. This allows to realize a situation where the renormalized tunneling amplitudes depend only weakly on the configuration $\hat{n}^{a,f}$ of the particles. For $x \approx x_{02} = 1.84$ the zeroth and second Bessel functions are approximately equal, $\mathcal{J}_0(x) \approx \mathcal{J}_2(x)$, and one can construct fully gauge invariant effective Hamiltonians. At $x \approx x_{01} = 1.43$ the zeroth and first Bessel functions are approximately equal.}
\label{figBessel}
\end{figure}

More generally, when the offset $\Delta_{2,1}= n \omega$ is a positive integer multiple $n=0,1,2,3,4,..$ of the driving frequency $\omega$, tunneling can also be restored. As shown by a general formalism in Ref.~\cite{Goldman2015}, the effective Hamiltonian in this case becomes
\begin{equation}
\H_{2,\rm eff} = - \tilde{t}_n \l  \ket{2}\bra{1} e^{i n \phi_{2,1}} + \ket{1}\bra{2} e^{-i n \phi_{2,1}}  \r.
\label{eqHeff2n}
\end{equation}
For $n=0$ the result is independent of the phase $\phi_{2,1}$ of the modulation. The tunneling matrix element is renormalized by
\begin{equation}
\tilde{t}_n =  t ~ \mathcal{J}_n \l \nicefrac{A_{2,1}}{\omega} \r.
\label{eqT0renormalization}
\end{equation}
The first three Bessel functions, $n=0,1,2$, are plotted in Fig.~\ref{figBessel} as a function of $x= A_{2,1} / \omega$.

Finally we consider the case when $\Delta_{2,1}= - n \omega$, for a positive integer $n=1,2,3,...$. In this case we can re-write the modulation \eqref{eqBondModToy} as
\begin{equation}
\Delta^\omega_{2,1}(t) = A_{2,1} \cos \l - \omega t - \phi_{2,1} \r,
\label{eqDltaOmgaNegw}
\end{equation}
i.e. effectively $\omega \to - \omega$ and $\phi_{2,1} \to - \phi_{2,1}$. By applying the results from Eq.~\eqref{eqHeff2n} and \eqref{eqT0renormalization} for the system with $- \omega$, we obtain
\begin{equation}
\H_{2,\rm eff} = - \tilde{t}_n \l  \ket{2}\bra{1} e^{i n ( \pi - \phi_{2,1} )} +  \ket{1}\bra{2} e^{- i n ( \pi - \phi_{2,1} )} \r.
\label{eqHeff2nNeg}
\end{equation}
The complex phase of the restored hopping in the effective Hamiltonian changes sign, because $- \phi_{2,1}$ appears in Eq.~\eqref{eqDltaOmgaNegw}. In addition, it contains a $\pi$ phase shift which takes into account the sign change of the renormalized tunneling matrix element $\propto \mathcal{J}_n(\nicefrac{A_{2,1}}{-\omega}) = e^{i \pi n} \mathcal{J}_n(\nicefrac{A_{2,1}}{\omega})$ if $n$ is odd.

\textbf{Multiple driving frequencies.}
Even more control over the restored tunnel couplings can be gained by using lattice modulations with multiple frequency components. Here we summarize results for the single-particle two-site problem from above, for the case of driving with frequency components $\omega$ and $2 \omega$. To do so, we modify our Hamiltonian in Eq.~\eqref{eqBondModToy} as
\begin{equation}
\H_2 = - t \l \ket{2}\bra{1} + \ket{1}\bra{2} \r + \Delta_{2,1}\ket{2}\bra{2} + \Delta^\omega_{2,1}(t) \ket{1}\bra{1},
\label{eqHbtoy2}
\end{equation}
where the $2 \pi/\omega$-periodic driving term takes the following form:
\begin{equation}
\Delta^\omega_{2,1}(t) = A_{2,1}^{(1)} \cos \l \omega t + \phi^{(1)}_{2,1} \r + A_{2,1}^{(2)} \cos \l 2 \omega t + \phi^{(2)}_{2,1} \r.
\label{eqBondModToyColor}
\end{equation}
In order to calculate the effective Hamiltonian, we rewrite the time-dependent Hamiltonian~\eqref{eqHbtoy2} in a moving frame, by performing a time-dependent unitary transformation realized by the operator~\cite{Goldman2015}
\begin{align}
\hat R (t)=& \exp \left ( i \Delta_{2,1} t \hat P_2  \right )  \exp \left \{i \l \frac{A_{2,1}^{(1)}}{\omega} \r \sin (\omega t + \phi^{(1)}_{2,1}) \hat P_1 \right \} \notag \\
& \times \exp \left \{i \l \frac{A_{2,1}^{(2)}}{2\omega} \r \sin (\omega t + \phi^{(2)}_{2,1}) \hat P_1 \right \},
\end{align}
where we introduced the projectors $\hat P_1\!=\!\vert  1 \rangle \langle 1 \vert$ and $\hat P_2\!=\!\vert  2 \rangle \langle 2 \vert$. In this moving frame, the time-dependent Hamiltonian in Eq.~\eqref{eqHbtoy2} takes the form
\begin{align}
\tilde{\mathcal{H}}_2 &= - t \ket{1}\bra{2} e^{- i \Delta_{2,1} t } \label{NewFrameHam}\\
& \times e^{i \l \frac{A_{2,1}^{(1)}}{\omega} \r \sin \left ( \omega t + \phi^{(1)}_{2,1} \right )} e^{i \l \frac{A_{2,1}^{(2)}}{2\omega} \r \sin \left (\omega t + \phi^{(2)}_{2,1} \right)} + \text{h.c.} \notag
\end{align}
Using the Jacobi-Anger identity,
\begin{equation}
e^{i \alpha \sin \left ( \omega t + \phi \right )} 
= \sum_{k=-\infty}^{\infty} \mathcal{J}_k \left (\alpha \right ) e^{i k \left (  \omega t + \phi \right ) },
\end{equation}
and time-averaging the time-dependent Hamiltonian in Eq.~\eqref{NewFrameHam} over a period $T\!=\!2 \pi/\omega$ of the drive, we obtain an effective Hamiltonian of the form 
\begin{equation}
\H_{2,\rm eff} = - \ket{1}\bra{2} \sum_{\ell = - \infty}^{\infty} \tilde{t}_{n,\ell} ~  e^{i (n - 2 \ell ) \phi_{2,1}^{(1)} + i \ell \phi_{2,1}^{(2)} } + \hc 
\label{eqHeffToySMcolor}
\end{equation}
While this effective Hamiltonian is similar to Eq.~\eqref{eqHeff2n}, the amplitude renormalization now involves a product of two Bessel functions:
\begin{equation}
\tilde{t}_{n,\ell} = t ~ \mathcal{J}_{n-2 \ell} \l \nicefrac{A_{2,1}^{(1)} }{\omega} \r ~ \mathcal{J}_{\ell} \l \nicefrac{A_{2,1}^{(2)} }{2 \omega} \r.
\label{eqT0renormalizationCol}
\end{equation}

\textbf{Two-particle two-site problem.}
Now we apply the results from the first paragraph [Eqs.~\eqref{eqHbtoy} - \eqref{eqHeff2nNeg}] to the problem of a pair of $a$ and $f$-particles in a double-well potential, see Fig.~\ref{figTwoWell}. In contrast to the main text, we consider general parameters in our derivation of the effective Hamiltonian. Our starting point is the model in Eqs.~\eqref{eqHintDef} - \eqref{eqHtComplete} for two sites $\vec{j}_1$ and $\vec{j}_2 = \vec{j}_1 +\vec{e}_y$. We assume $V_a(\vec{j}_{1,2}) \equiv 0$ but introduce a static energy offset $\Delta^f = U$ between the two lattice sites for the $f$-particles, $V_f(\vec{j}_2) = \Delta^f + V_f(\vec{j}_1)$. Because our analysis is restricted to the subspace with one $a$-particle and one $f$-particle, the hard-core constraint assumed in the main text is not required in this case and the statistics of the two species are irrelevant. 

\begin{figure}[t!]
\centering
\epsfig{file=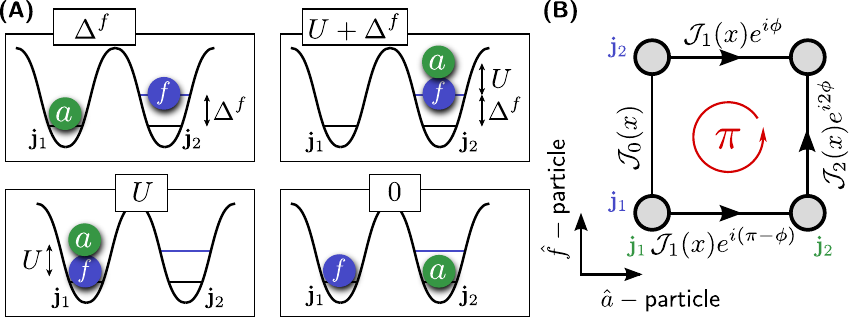, width=0.5\textwidth}
\caption{\textbf{Two-site two-particle problem.} (A) We consider one $a$ and one $f$-type particle tunneling between the sites $\vec{j}_{1,2}$ of a double-well potential. The strong potential gradient $\Delta^f$, seen by $f$-particles, and inter-species Hubbard interactions $U$ suppress coherent tunnelings $t_y^a$ and $t_y^f$ of $a$ and $f$-particles, respectively. (B) As described in the text, the resonant lattice modulation $\H_\omega(t)$ with frequency $\omega = U = \Delta^f$ restores the tunnel couplings between the four two-particle basis states, with amplitudes and phases indicated in the figure, where $\phi = \phi_{\vec{j}_2,\vec{j}_1}$. This induces a $\pi$-flux in the plaquettes of the many-body Hilbert space, allowing to implement $\Zt$ LGT.}
\label{fig2P2S}
\end{figure}

The two-site problem has four basis states, $\fd_{\vec{j}_{m}} \ad_{\vec{j}_{n}} \ket{0}$ with $m,n=1,2$. Their corresponding on-site energies are $0$, $\Delta^f=U$, $U$, $\Delta^f+U=2 U$, see Fig.~\ref{fig2P2S} (A), which suppress most coherent tunneling processes because $\Delta^f = U \gg |t_y^a|, |t_y^f|$. When the resonant lattice modulation $\H_\omega(t)$ with frequency $\omega = U$ is included, all tunnel couplings are restored. Now we will show that the effective Floquet Hamiltonian is given by
\begin{equation}
\H_{\rm eff}^{\rm 2well} = - t_y^a ~ \lambda ~ e^{i \hat{\varphi}} ~ \ad_{\vec{j}_2} \a_{\vec{j}_1} - t_y^f ~  \hat{\Lambda} ~ e^{i \hat{\theta}} ~ \hat{\tau}^+_{\langle \vec{j}_2,\vec{j}_1 \rangle} + \hc~,
\label{eqHeff2s2pComb}
\end{equation}
where $\hat{\tau}^+_{\langle \vec{j}_2,\vec{j}_1 \rangle} = \fd_{\vec{j}_2} \f_{\vec{j}_1}$ and
\begin{flalign}
\hat{\varphi} &= \phi_{\vec{j}_2,\vec{j}_1} + \l 1 - \hat{\tau}^z_{\langle \vec{j}_2,\vec{j}_1 \rangle} \r   \l \frac{\pi}{2} - \phi_{\vec{j}_2,\vec{j}_1}  \r,  \label{eqHeff2s2pPhi}\\
\lambda &= \mathcal{J}_1 \l A_{\vec{j}_2,\vec{j}_1} / \omega \r, \label{eqHeff2s2plbda} \\
\hat{\theta} &= 2 \phi_{\vec{j}_2,\vec{j}_1}  \hat{n}^a_{\vec{j}_2}, \label{eqHeff2s2pTht} \\
\hat{\Lambda} &= \mathcal{J}_0 \l \nicefrac{A_{\vec{j}_2,\vec{j}_1}}{\omega} \r \hat{n}^a_{\vec{j}_1} + \mathcal{J}_2 \l \nicefrac{A_{\vec{j}_2,\vec{j}_1}}{\omega} \r \hat{n}^a_{\vec{j}_2}. \label{eqHeff2s2pLbda}
\end{flalign}

To derive Eqs.~\eqref{eqHeff2s2pComb} - \eqref{eqHeff2s2pLbda}, we first consider the effect of the coherent driving $\H_\omega(t)$, characterized by Eq.~\eqref{eqDefVomega}, on the matter field $\a$. Because the Hamiltonian 
\begin{multline}
\H^a = - t_y^a \l \ad_{\vec{j}_2} \a_{\vec{j}_1} + \hc \r + \l \hat{n}^a_{\vec{j}_2} - \hat{n}^a_{\vec{j}_1} \r \\ 
\times \frac{1}{2} \l  U \hat{\tau}^z_{\langle \vec{j}_2, \vec{j}_1 \rangle} +  V_\omega(\vec{j}_2,t) - V_\omega(\vec{j}_1,t) \r, 
\end{multline}
governing the dynamics of $\a$, commutes with the link variable, characterizing the gauge field, $[ \H^a , \hat{\tau}^z_{\langle \vec{j}_2, \vec{j}_1 \rangle}] = 0$, we can treat $\hat{\tau}^z_{\langle \vec{j}_2, \vec{j}_1 \rangle}$ as a $\mathbb{C}$-number with two possible values, $\pm 1$. 

When expressed in terms of the two states $\ket{1}_a = \ad_{\vec{j}_1} \ket{0}$ and $\ket{2}_a = \ad_{\vec{j}_2} \ket{0}$, the Hamiltonian $\H^a$ is of the same form as $\H_2$ in Eq.~\eqref{eqHbtoy}. It has an energy difference of $\Delta_{2,1} = U \hat{\tau}^z_{\langle \vec{j}_2, \vec{j}_1 \rangle}$ between the two states, which is caused microscopically by the inter-species Hubbard interaction $U$, see Fig.~\ref{fig2P2S} (A).

According to Eqs.~\eqref{eqHeff2n}, \eqref{eqHeff2nNeg}, the restored tunnel coupling between $\ket{1}_a$ and $\ket{2}_a$ has a complex phase given by $\varphi = \phi_{\vec{j}_2,\vec{j}_1}$ if $\Delta_{2,1} > 0$, i.e. for $\tau^z_{\langle \vec{j}_2, \vec{j}_1 \rangle} = 1$, and it is $\varphi = \pi -\phi_{\vec{j}_2,\vec{j}_1}$ if $\Delta_{2,1} < 0$, i.e. for $\tau^z_{\langle \vec{j}_2, \vec{j}_1 \rangle} = -1$. Because in both cases the magnitude of the energy mismatch between the two sites is $|\Delta_{2,1}| = \omega$, the tunneling is renormalized by $\lambda =  \mathcal{J}_1 \l A_{\vec{j}_2,\vec{j}_1} / \omega \r$. These results confirm Eqs.~\eqref{eqHeff2s2pPhi}, \eqref{eqHeff2s2plbda}.

Next we consider the dynamics of the $f$-particles, or, equivalently, the link variable $\hat{\tau}^z_{\langle \vec{j}_2, \vec{j}_1 \rangle}$. It is governed by the following Hamiltonian
\begin{multline}
\H^f = - t_y^f ~ \l \hat{\tau}^+_{\langle \vec{j}_2,\vec{j}_1 \rangle} + \hc \r + \hat{\tau}^z_{\langle \vec{j}_2,\vec{j}_1 \rangle} \\
\times \frac{1}{2} \l \Delta^f + U \delta \hat{n}^a + V_\omega(\vec{j}_2,t) - V_\omega(\vec{j}_1,t) \r .
\end{multline}
Because $\H^f$ commutes with the matter field, $[\H^f, \hat{n}^a_{\vec{j}_{1}}] = [\H^f, \hat{n}^a_{\vec{j}_{2}}]  = 0$, we can treat the particle number imbalance 
\begin{equation}
\delta \hat{n}^a = \hat{n}^a_{\vec{j}_2} -  \hat{n}^a_{\vec{j}_1}
\end{equation}
as a $\mathbb{C}$-number now, which can take two  values $\pm 1$. 

When expressed in terms of the two states $\ket{1}_f = \fd_{\vec{j}_1} \ket{0}$ and $\ket{2}_f = \fd_{\vec{j}_2} \ket{0}$, the Hamiltonian $\H^f$ is of the same form as $\H_2$ in Eq.~\eqref{eqHbtoy}. It has an energy difference of $\Delta_{2,1} = \Delta^f + U \delta \hat{n}^a$ between the two states, which is caused microscopically by the inter-species Hubbard interaction $U$ and the potential gradient $\Delta^f$ which the $f$-particles are subject to, see Fig.~\ref{fig2P2S} (A).

In the case of $f$-particles, the energy offset $\Delta_{2,1}$ can only take positive values $0$ and $2 \omega$ if $\Delta^f = U = \omega$. From Eq.~\eqref{eqHeff2n} it follows that the restored tunnel coupling between $\ket{1}_f$ and $\ket{2}_f$ has a complex phase given by $\theta = 0$ if $\Delta_{2,1}=0$, i.e. for $\delta n^a = -1$, and by $\theta = 2 \phi_{\vec{j}_2,\vec{j}_1}$ if $\Delta_{2,1}=2 \omega$, i.e. for $\delta n^a = 1$. Expressed in terms of $\hat{n}^a_{\vec{j}_2}$, in a subspace where $\hat{n}^a_{\vec{j}_1} + \hat{n}^a_{\vec{j}_2} = 1$, this result confirms Eq.~\eqref{eqHeff2s2pTht}.

The magnitudes of the restored tunneling couplings of $f$-particles in the two-particle Hilbert space depend on the energy offset $\Delta_{2,1}$. In the case when $\Delta_{2,1}=0$, i.e. for $\delta n^a = -1$, it becomes $\Lambda t_y^f = t_y^f \mathcal{J}_0( A_{\vec{j}_2,\vec{j}_1} / \omega)$. When $\Delta_{2,1}=2 \omega$, i.e. for $\delta n^a = 1$, it is given another Bessel function, $\Lambda t_y^f = t_y^f  \mathcal{J}_2( A_{\vec{j}_2,\vec{j}_1} / \omega)$. This result, summarized in Fig.~\ref{fig2P2S} (B), confirms Eq.~\eqref{eqHeff2s2pLbda}.

\begin{figure*}[t!]
\centering
\epsfig{file=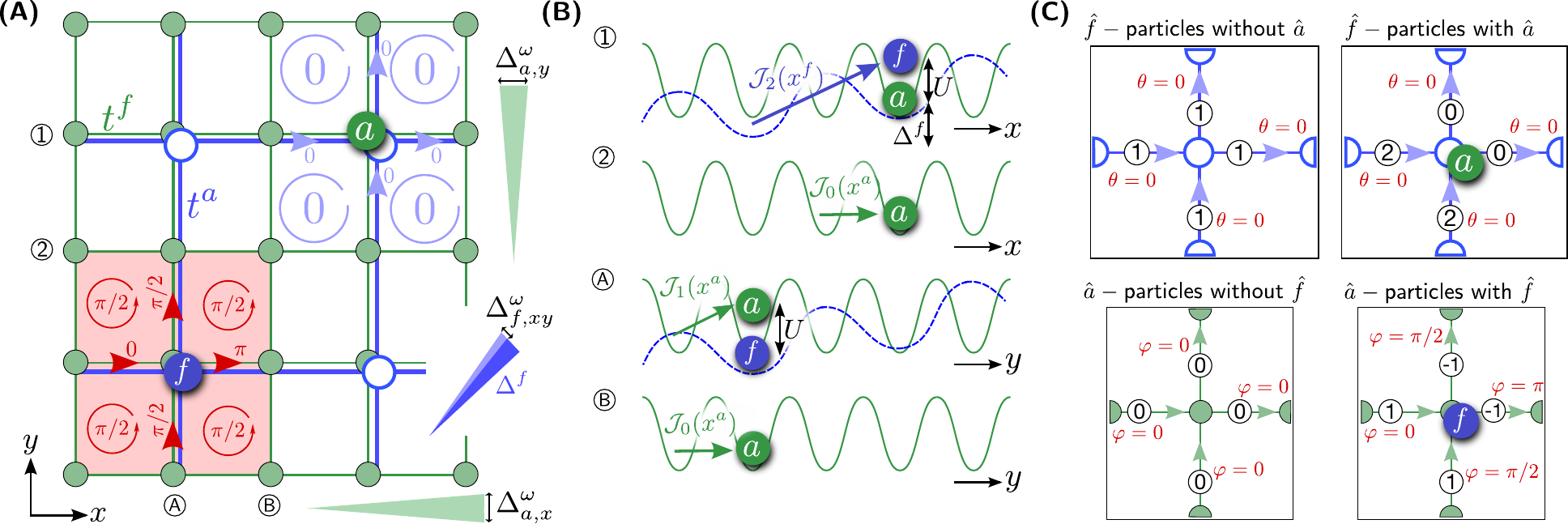, width=0.99\textwidth}
\caption{\textbf{Flux-attachment in a 2D Hofstadter model.} (A) We consider a two-component mixture of $a$-particles tunneling on a short square lattice (green) and $f$-type particles hopping between the sites of a long square lattice (blue). These tunnel couplings are first suppressed by strong inter-species Hubbard interactions, and for $f$-particles by an additional static potential gradient $\Delta^f_{xy}$ along $\vec{e}_x + \vec{e}_y$. (B) Using resonant lattice modulations $\Delta^\omega_{a,\mu}$ and $\Delta^\omega_{f,xy}$, tunnel couplings are restored. As a result, the matter field $\a$ is subject to synthetic magnetic fluxes localized around the $f$-particles, as illustrated in the bottom left corner in (A). The $f$-particles are not subject to any synthetic gauge fields. (C) The complex phases and renormalized amplitudes of the restored tunnelings can be inferred by calculating the energy offsets $\Delta = n \omega$ involved in the elementary hopping processes. The results in units of the driving frequency $\omega$, $n$, are indicated in circles on the respective bonds. If $n \geq 0$, the restored phase on bond $\ij$ is $n \phi_{\ij}$, where $\phi_{\ij}$ is the phase of the relative modulation between sites $\vec{i}$ and $\vec{j}$; If $n < 0$, the restored phase is $|n| ( \pi - \phi_{\ij} )$. The amplitude renormalization is given by $\mathcal{J}_{|n|}(A_{\ij} / \omega)$, where $A_{\ij}$ is the amplitude of the relative modulation between sites $\vec{i}$ and $\vec{j}$.}
\label{fig2DFluxAttachment}
\end{figure*}

\textbf{Realizations with ultracold atoms.}
Next we discuss realizations of the two-particle two-site problem with ultracold atoms. The proposed implementation needs two distinguishable particles with strong inter-species on-site interaction energy $U\gg t_{y}$. The particles occupy a double well with both species-dependent and species-independent on-site potentials. For the species-dependent contribution a static potential is sufficient, which introduces a tilt $\Delta^{f}=U$ between neighboring sites for the $f$-particles but leads to zero tilt for the $a$-particles. On the other hand, the species-independent contribution must be time-dependent $V_{\omega}(t)$, in order to restore resonant tunneling for both particles.

For ultracold atoms a cubic array of lattice sites with period~$d_\mathrm{s}$ can be created by three mutually orthogonal standing waves with wavelengths $\lambda=2d_\mathrm{s}$. When extending this simple cubic lattice along one axes by an additional lattice with twice the period $d_\mathrm{l}=2d_\mathrm{s}$, a superlattice of the form $V_\mathrm{s}\sin^2(\pi y/d_\mathrm{s} + \pi/2) + V_\mathrm{l}\sin^2(\pi y/d_\mathrm{l} + \phi_\mathrm{SL}/2)$ arises. In the limit $V_\mathrm{l}\gg V_\mathrm{s}$ the superlattice potential resembles a chain of double wells, where tunneling between each double well is suppressed and all dynamics is restricted to two sites. Tuning the relative phase~$\phi_\mathrm{SL}$ allows for dynamic control of the on-site potentials. In principle the time-dependent modulation~$V_{\omega}(t)$ can be implemented by a fast modulation of~$\phi_\mathrm{SL}$; however, the modulation frequency may be limited to small values depending on the implementation of the lattices. For a superlattice with a common retro-reflector for instance, the phase $\phi_\mathrm{SL}$ can only be varied by changing the frequency of the laser. Alternatively, a second lattice $V_\mathrm{mod}\sin^2(\pi y/d_\mathrm{l} + \phi_\mathrm{mod}/2)$ with period~$d_\mathrm{l}$ and phase~$\phi_\mathrm{mod}=\pm\pi/2$ can be introduced, such that it only affects the on-site potential of a single site of the double well. Therefore, amplitude modulation~$V_\mathrm{mod}(t)$ of this additional lattice induces a relative modulation of the on-site energies. This leads to a non-zero species-independent, time averaged energy offset, which can be compensated by the static phase degree of freedom~$\phi_\mathrm{SL}$ of the superlattice.

The two distinguishable particles can be encoded in different hyperfine sublevels with different magnetic moments, enabling the direct implementation of the static species-dependent potentials by a magnetic field gradient. This is especially appealing for bosonic atoms possessing a hyperfine sublevel with zero magnetic moment, which directly results in a vanishing, magnetic field independent tilt for the $a$-particles in first order. Nevertheless, this is not essential as tilts for the $a$-particles can be compensated by the present species-independent potentials.

\noindent
\textbf{II. Flux-attachment in 2D}\\
Here we discuss a situation where the $f$-particles become sources of magnetic flux for $a$-particles in a 2D Hofstadter model, as illustrated in Fig.~\ref{figSetup} (A) of the main text. Specifically we demonstrate how superlattices can be used to realize the case when the magnetic flux in all plaquettes including the $f$-particle becomes $\Phi=\pi/2$, i.e. the $f$-particle is bound to exactly one flux quantum seen by the matter field $\a$.

We propose a setup as sketched in Fig.~\ref{fig2DFluxAttachment} (A). The $f$-particles are tunneling between the sites of a long square lattice, with bare tunneling amplitude $t^f$. In addition, they are subject to a potential gradient 
\begin{equation}
V_f(\vec{j}) = \vec{j} \cdot \l \vec{e}_x + \vec{e}_y \r \frac{\Delta^f}{4},
\label{eqVfxyStat}
\end{equation}
where $\vec{j} = (2 n^x, 2n^y)$ for $n^{x,y} \in \mathbb{Z}$ denotes the sites of the long lattice. As before we assume that the $f$-particles are hard-core bosons, and we note that a generalization to fermions is straightforward. 

The particles $a$, describing the matter field, tunnel between the sites $\vec{i} = (n^x, n^y)$, for $n^{x,y} \in \mathbb{Z}$, of the short square lattice, with amplitude $t^a$. They interact with the $f$-particles by local Hubbard interactions $U$, and we assume that they are hard-core bosons; again, a generalization to fermions is straightforward. For the $a$-particles, no external potential is required, i.e. we consider $V_a(\vec{i}) \equiv 0$. 

To restore the tunnel couplings, which are suppressed by $\Delta^f = U = \omega$, we consider a state-dependent driving term 
\begin{equation}
\H_\omega = \sum_{\vec{j}} V^\omega_f(\vec{j},t) \fd_{\vec{j}} \f_{\vec{j}} + \sum_{\vec{i}} V^\omega_a(\vec{i},t) \ad_{\vec{i}} \a_{\vec{i}}.
\end{equation}
Note that this is different from the situation discussed in the main text, where both species are subject to the same driving. However the lattice modulation we consider here is particularly easy to implement: To restore tunneling of the matter field, we require oscillating potential gradients $\Delta^\omega_{a,\mu}$ along $\mu=x,y$ directions,
\begin{equation}
V^\omega_a(\vec{i},t) = \sum_{\mu =x,y} \Delta^\omega_{a,\mu} ~ \vec{i} \cdot \vec{e}_\mu \cos \l \omega t + \phi_a^\mu \r. 
\end{equation}
Similarly, the potential gradient in Eq.~\eqref{eqVfxyStat} is modulated in order to restore tunneling of the $f$-particles,
\begin{equation}
V^\omega_f(\vec{j},t) = \frac{ \Delta^\omega_{f,xy}}{4}   ~\vec{j} \cdot \l \vec{e}_x + \vec{e}_y \r  \cos \l \omega t + \phi_f \r.
\label{eqVfxyDyn}
\end{equation}

In the following we consider the phase choice,
\begin{equation}
\phi_a^y = \pi/2, \qquad \phi_a^x = 0, \qquad \phi_f = 0.
\end{equation}
Moreover we assume that the dimensionless driving strengths are given by
\begin{equation}
\Delta^\omega_{a,\mu} / \omega = x_{01} \approx 1.43, \qquad \Delta^\omega_{f,xy} / \omega = x_{02} \approx 1.84.
\label{eq2DFluxDrvgStrgth}
\end{equation}
For these values it holds $\mathcal{J}_0(x_{01}) = \mathcal{J}_1(x_{01})$ and $\mathcal{J}_0(x_{02}) = \mathcal{J}_2(x_{02})$, see Fig.~\ref{figBessel}.

As described above Eq.~\eqref{eqHeffGeneral} in the main text, when $\omega \gg t^a, t^f$ the system can be described by an effective hopping Hamiltonian. For the setup described here, we obtain the following result,
\begin{multline}
\H_{\rm eff} = - t^f \sum_{\langle \vec{j}_2,\vec{j}_1 \rangle} \l  \fd_{\vec{j}_2} \f_{\vec{j}_1} \hat{\Lambda}_{\langle \vec{j}_2,\vec{j}_1 \rangle} + \hc \r  \\
-  t^a \sum_{\langle \vec{i}_2,\vec{i}_1 \rangle} \l  \ad_{\vec{i}_2} \a_{\vec{i}_1} \hat{\lambda}_{\langle \vec{i}_2,\vec{i}_1 \rangle} e^{i \hat{\varphi}_{\langle \vec{i}_2,\vec{i}_1 \rangle}} + \hc \r.
\label{eqHeff2DfluxAttchment}
\end{multline}
Expressions for the amplitudes and phases can be derived from the energy offsets, as sketched in Fig.~\ref{fig2DFluxAttachment} (C). This leads to the following results:

(i) The complex phase for tunneling of $f$-particles is always zero, and their amplitude renormalization is given by the expression
\begin{multline}
 \hat{\Lambda}_{\langle \vec{j}_2,\vec{j}_1 \rangle} = \mathcal{J}_1\l x_{02} \r \l 1 - (\delta \hat{n}^a_{\langle \vec{j}_2,\vec{j}_1 \rangle})^2 \r \\
 + \mathcal{J}_0\l x_{02} \r  (\delta \hat{n}^a_{\langle \vec{j}_2,\vec{j}_1 \rangle})^2,
\end{multline}
where $\delta \hat{n}^a_{\langle \vec{j}_2,\vec{j}_1 \rangle}$ denotes the imbalance of the matter field on the bond $\langle \vec{j}_2,\vec{j}_1 \rangle$, i.e.
\begin{equation}
\delta \hat{n}^a_{\langle \vec{j}_2,\vec{j}_1 \rangle} = \hat{n}^a_{\vec{j}_2} - \hat{n}^a_{\vec{j}_1}.
\end{equation}

(ii) The complex phase for tunneling of $a$-particles depends on the density $\hat{n}^f$ of $f$-particles, as illustrated in Fig.~\ref{fig2DFluxAttachment} (A) in the left, lower corner. Written out explicitly, we obtain the following expression,
\begin{flalign}
\hat{\varphi}_{\langle \vec{j},\vec{j}-\vec{e}_x \rangle} &= 0, \\
\hat{\varphi}_{\langle \vec{j},\vec{j}-\vec{e}_y \rangle} &= \hat{n}^f_{\vec{j}} ~ \pi/2, \\
\hat{\varphi}_{\langle \vec{j} + \vec{e}_x,\vec{j} \rangle} &=  \hat{n}^f_{\vec{j}} ~ \pi, \\
\hat{\varphi}_{\langle \vec{j}+\vec{e}_y,\vec{j} \rangle} &=  \hat{n}^f_{\vec{j}} ~ \pi/2.
\end{flalign}
Note that $\vec{j}$ corresponds to a site from the long lattice in these expressions. In the remaining cases, not included above, $\hat{\varphi}_{\langle \vec{i}_2,\vec{i}_1 \rangle}=0$. For the chosen driving strength, see Eq.~\eqref{eq2DFluxDrvgStrgth}, the amplitude renormalization becomes $\hat{\lambda}_{\langle \vec{i}_2,\vec{i}_1 \rangle} = \mathcal{J}_0(x_{01}) \approx 0.55$ on all bonds. 

Eq.~\eqref{eqHeff2DfluxAttchment} realizes a situation where $f$-particles carry one unit of magnetic flux, seen by the matter field. As claimed in the main text, this situation can be implemented using the general scheme proposed in this work. By varying the driving frequency and the phases of the lattice modulation, many more interesting Hamiltonians can be realized with our scheme. An interesting example corresponds to the choice $\mathcal{J}_{0}( \nicefrac{\Delta^\omega_{a,y} }{ \omega}) = 0$, for which $a$-particles can only move in the presence of $f$-particles.

~ \\
\textbf{III. Implementing matter coupled to a $\Zt$ gauge field in the two-leg ladder geometry}\\
Here we describe in detail our implementation scheme how the matter field can be coupled to a $\Zt$ lattice gauge field in a two-leg ladder geometry. We start by defining the model parameters of the considered system and specify all terms in the general Hamiltonian \eqref{eqHtComplete}. The setup is summarized in Fig.~\ref{figTwoLegLadder} of the main text.

\textbf{Model.} 
We consider a situation where every rung $(j \vec{e}_x, j \vec{e}_x+\vec{e}_y)$ of the ladder is occupied by exactly one $f$-particle, which requires $t^f_x=0$. This can be achieved directly by a deep state-dependent lattice, or, alternatively, using a very strong gradient $|\Delta_x^f| \gg U$ to suppress the bare tunneling $t^f_x$. When $\Delta_x^f = n U$ is a large integer multiple $n \gg 1$ of the resonant modulation frequency $\omega = U$, the amplitude of the restored tunneling is renormalized by $\mathcal{J}_n(x)$ which decays exponentially with $n$. For example, for the driving strength $x_{02} = 1.84$ discussed in the main text, $\mathcal{J}_5(x_{02}) = 4.8 \times 10^{-3}$, and already for $n=10$ we have $\mathcal{J}_{10}(x_{02}) = 1.1 \times 10^{-7}$ which is practically zero. Since every rung is occupied by just one $f$-particle, the statistics of the latter become irrelevant and no hard-core constraint is required. 

As in the two-site problem the tunneling $t^f_y$ is freely tunable and we require a linear potential gradient $\Delta^f_y = U = \omega$ along the rungs of the ladder; the latter is resonant with the lattice modulation frequency $\omega$. Summarizing, we consider the following parameters,
\begin{equation}
V_f(\vec{j}) = j_x \Delta^f_x + j_y U, \quad t^f_x=0.
\end{equation}

The $a$-particles are free to move along both directions of the ladder, with bare tunnel couplings $t^a_{x,y}$. We assume that they are hard-core bosons. This can be realized experimentally by very strong intra-species Hubbard interactions
\begin{equation}
\H_{\rm int}^a = \frac{U_a}{2} \sum_{\vec{j}} \hat{n}^a_{\vec{j}} ( \hat{n}^a_{\vec{j}} - 1),
\end{equation}
with $U_a \gg U$. When $U_a = n U$ is a large integer multiple $n \gg 1$ of the inter-species interactions $U=\omega$, the tunneling matrix elements from singly occupied states into states with two $a$ bosons on one site are strongly suppressed. As discussed above for $f$-particles, the lattice modulation in Eq.~\eqref{eqHomegaDef} leads to tunneling amplitudes renormalized by the Bessel functions $\mathcal{J}_n(x)$. They are exponentially small for large $n$ and can thus be neglected in the effective Hamiltonian. 

To control the phases $\hat{\varphi}^x$ of the tunnel couplings for $a$-particles along the legs of the ladder, restored by the lattice modulation \eqref{eqHomegaDef}, we consider an additional gradient $\Delta^a_x = U = \omega$ in $x$-direction,
\begin{equation}
V_a(\vec{j}) = \Delta^a_x ~ j_x, \qquad \Delta^a_x=\omega.
\end{equation}
The number of $a$-particles $N_a$ in the system is freely tunable from $N_a=0$ to $N_a = 2 L_x$.

Finally, the lattice modulation \eqref{eqHomegaDef} is defined by a time-dependent gradient seen equally by both species,
\begin{equation}
V_\omega(\vec{j},t) =  (V_\omega^x  j_x +V_\omega^y   j_y ) \cos(\omega t + \phi), \quad \phi = 0.
\end{equation}
In the following derivation of the effective Floquet Hamiltonian we consider arbitrary amplitudes $V_\omega^{x,y}$, but we assume that the phase of the modulation is trivial, $\phi=0$. Later we will show that the model has a $\Zt$ gauge structure for the specific choice $V_\omega^{x}/\omega = V_\omega^{y}/\omega = x_{02}$.

\begin{figure}[t!]
\centering
\epsfig{file=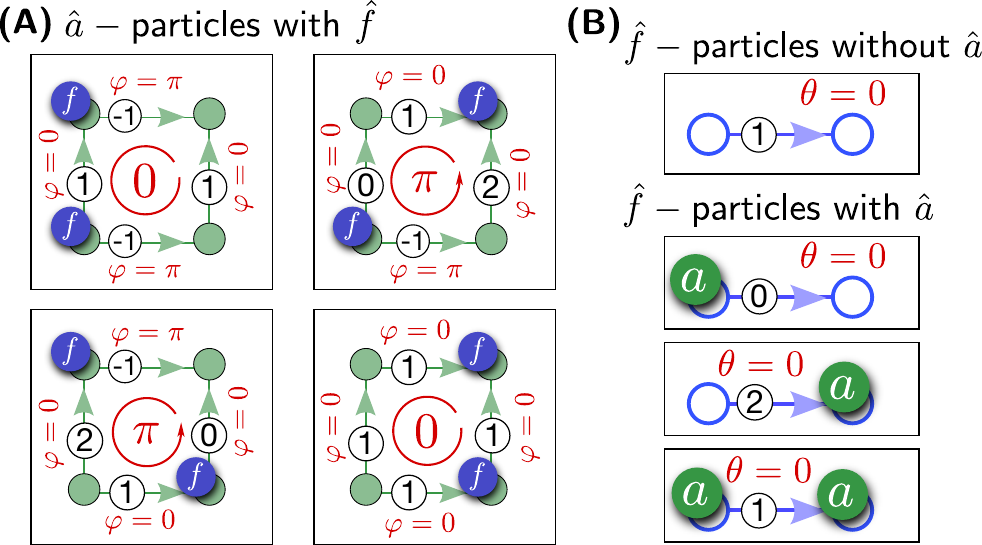, width=0.48\textwidth}
\caption{\textbf{Derivation of the effective Hamiltonian.} We consider all possible hopping processes of $a$-particles in the presence of different $f$ configurations (A), and all hopping processes of $f$-particles in the presence of different $a$ configurations (B). The numbers $n$ in circles indicated the energy difference in units of $\omega$ that is restored by the lattice modulation when a particle tunnels on the respective bond in positive $\vec{e}_x$ or $\vec{e}_y$ direction. The phases $\phi$ and $\theta$ of the respective tunnelings are indicated, and the amplitude renormalization is given by $\mathcal{J}_{|n|}(x)$, where $x$ is the dimensionless driving strength on the corresponding bond.}
\label{figZ2LGTderivation}
\end{figure}

\textbf{Effective Hamiltonian.} 
Now we apply the general formalism discussed above and derive the effective Hamiltonian $\H_{\rm eff}$ of the form Eq.~\eqref{eqHeffGeneral} which governs the dynamics of the system in the strong driving limit $\omega \gg |t^\alpha_\mu|$ for $\alpha=a,f$ and $\mu=x,y$. To derive the phases, $\hat{\varphi}$ and $\hat{\theta}$, and amplitudes, $\hat{\lambda}$ and $\hat{\Lambda}$, of the restored tunneling matrix elements, we consider all possible hopping processes involving one, two or three particles. 

By going through all cases, see Fig.~\ref{figZ2LGTderivation}, we arrive at the following expressions for the phases,
\begin{flalign}
\hat{\varphi}^y_{\ij} &=  \pi ( 1 - \hat{\tau}^z_{\ij} ) / 2, \qquad \hat{\varphi}^x_{\ij} = 0,  \label{eqDervtnZ2varphi}\\
\hat{\theta}^y_{\ij} &= 0.
\end{flalign}
For the amplitude renormalizations we obtain
\begin{equation}
\hat{\lambda}^y = \mathcal{J}_1 \l \frac{V_\omega^y}{\omega} \r,
\end{equation}
and assuming that $\vec{i} = \vec{j} + \vec{e}_x$, we get
\begin{multline}
\hat{\lambda}_{\ij}^x = \frac{1}{2} \l 1 + \hat{\tau}_{\langle \vec{j}, \vec{l} \rangle}^z \hat{\tau}_{\langle \vec{i}, \vec{k} \rangle}^z \r  \mathcal{J}_1 \l \frac{V_\omega^x}{\omega} \r +   \frac{1}{4} \times \\
 \bigg\{ \left[ 1 + (-1)^{j_y} \hat{\tau}_{\langle \vec{j}, \vec{l} \rangle}^z \right] \left[ 1 -  (-1)^{j_y}  \hat{\tau}_{\langle \vec{i}, \vec{k} \rangle}^z \right] \mathcal{J}_2 \l \frac{V_\omega^x}{\omega} \r \\
+ \left[ 1 - (-1)^{j_y} \hat{\tau}_{\langle \vec{j}, \vec{l} \rangle}^z \right] \left[ 1 +  (-1)^{j_y}  \hat{\tau}_{\langle \vec{i}, \vec{k} \rangle}^z \right] \mathcal{J}_0 \l \frac{V_\omega^x}{\omega} \r \bigg\}.
\label{eqlambdajmux}
\end{multline}
Here $\vec{l} = \vec{j} \pm \vec{e}_y$ and $\vec{k} = \vec{i} \pm \vec{e}_y$ denote the lattice sites on the opposite ends of the rungs including $\vec{j}$ and $\vec{i}$. For the calculation of the amplitude renormalization for $f$-particles, we assume that $\vec{i} = \vec{j} + \vec{e}_y$ and obtain
\begin{multline}
\hat{\Lambda}^y_{\ij} =  \frac{1}{4} \bigg\{  \left[ 1 + (-1)^{\hat{n}^a_{\vec{j}}} \right] \left[ 1 - (-1)^{\hat{n}^a_{\vec{i}}}  \right] \mathcal{J}_2 \l \frac{V_\omega^y}{\omega} \r \\
+   \left[ 1 - (-1)^{\hat{n}^a_{\vec{j}}}  \right] \left[ 1 + (-1)^{\hat{n}^a_{\vec{i}}}  \right] \mathcal{J}_0 \l \frac{V_\omega^y}{\omega} \r  \bigg\} \\
+ \frac{1}{2} \l 1 + (-1)^{\hat{n}^a_{\vec{j}} + \hat{n}^a_{\vec{i}}} \r  \mathcal{J}_1 \l \frac{V_\omega^y}{\omega} \r.
\label{eqLambdaxj}
\end{multline}

In the main text we discuss the effective Hamiltonian for specific values of the driving strengths, 
\begin{equation}
V_\omega^x / \omega = V_\omega^y / \omega = x_{02} \approx 1.84.
\label{eqDefDrvgStrngthsLdr}
\end{equation}
Because $\mathcal{J}_0(x_{02}) = \mathcal{J}_2(x_{02})$ the expressions \eqref{eqlambdajmux}, \eqref{eqLambdaxj} simplify significantly. The amplitude renormalization $\hat{\lambda}_{\ij}^y$ depends only on the product $\hat{\tau}_{\langle \vec{j}, \vec{l} \rangle}^z \hat{\tau}_{\langle \vec{i}, \vec{k} \rangle}^z$, and $\hat{\Lambda}$ depends only on the total number of $a$-particles, modulo two, on the rung, $(-1)^{\hat{n}^a_{\vec{j}} + \hat{n}^a_{\vec{i}}}$. Because the phase $\hat{\varphi}^y$ takes values $0$ or $\pi$, see Eq.~\eqref{eqDervtnZ2varphi}, it follows that
\begin{equation}
e^{i \hat{\varphi}^y_{\ij}} =  \hat{\tau}^z_{\ij} = \pm 1.
\end{equation}

For the choice of the driving strengths in Eq.~\eqref{eqDefDrvgStrngthsLdr}, the effective Hamiltonian Eq.~\eqref{eqHeffLadder} of the main text is obtained. Written in terms of the plaquette operators from Eq.~\eqref{eqDefBp}, the amplitude renormalizations in Eqs.~\eqref{eqlambdajmux}, \eqref{eqLambdaxj} become
\begin{multline}
\hat{\lambda}^x_{\ij_x} = \frac{1}{2} \l 1 - \hat{B}_{p(\ij_x)} \r  \mathcal{J}_0(x_{02})\\
+ \frac{1}{2} \l 1 +  \hat{B}_{p(\ij_x)} \r  \mathcal{J}_1(x_{02})
\label{eqDeflbdax}
\end{multline}
for $a$-particles, where $p(\ij_x)$ denotes the plaquette in the ladder which includes bond $\ij_x$; in Eq.~\eqref{eqDeflbdaxMain} of the main text, $\hat{B}_{p(\ij_x)}$ is written out explicitly in terms of the $\Zt$ gauge field $\hat{\tau}^z$ on the rungs. For $f$-particles,
\begin{multline}
\hat{\Lambda}^y_{\ij_y} = \frac{1}{2} \l 1 - \hat{Q}_{\vec{i}} \hat{Q}_{\vec{j}} \r \mathcal{J}_0(x_{02})\\
+ \frac{1}{2} \l 1 + \hat{Q}_{\vec{i}} \hat{Q}_{\vec{j}}   \r  \mathcal{J}_1(x_{02}).
\label{eqDefLBDA}
\end{multline}

\textbf{Realizations with ultracold atoms.}
Here we want to sketch a realization of the presented ladder model with ultracold atoms. The proposed realization is a direct extension of the two-site two-particle problem discussed before, by coupling the double well potentials for the $a$-particles along the orthogonal $x$-direction. Note, it is essential that tunneling for the $f$-particles is suppressed. To this end, either the tunneling rates $t_x^\mu$ itself can be different by using a species-dependent lattice along~$x$, or the sites can be energetically detuned for the $f$-particles by applying a species-dependent gradient potential $\Delta_x^f\gg t_x^f$. To engineer the appropriate complex tunneling matrix elements for $a$-particles, neighboring sites along~$x$ need to be tilted by~$\Delta_x^a=U$ and modulated in time using dynamic gradient potentials.

 ~\\
\textbf{IV. Gauge structure of two-leg ladders}\\
In this section we study the $\Zt$ gauge structure of the effective Hamiltonians $\H_{\rm 2leg}$, $\H_{\rm 2leg}^{\rm simp}$ in the two-leg ladder geometry, see Eqs.~\eqref{eqHeffLadder}, \eqref{eqDefHLGT}. To this end we introduce generalizations of these models which have local $\Zt$ gauge invariance on all sites. Using the approach introduced at the end of the main text, these extended models can also be directly realized experimentally. We will demonstrate that their phase diagram is identical to that of the simpler models Eq.~\eqref{eqHeffLadder} and \eqref{eqDefHLGT}, although the local order parameters of the latter have to be replaced by non-local gauge-invariant versions in the extended models which we derive here.

\textbf{Fully $\Zt$ gauge invariant formulation.} 
Now we introduce an extended Hilbert space and introduce additional link variables $\tau^z_{\ij_x}$ on the legs of the ladder.

\emph{Effective Hamiltonian.--}
The original Hamiltonian \eqref{eqHeffLadder} is obtained from the more general one,
\begin{multline}
\tilde{\mathcal{H}}_{\rm 2leg} = - \sum_{\ij_x} \l  t^a_x ~ \hat{\lambda}^x_{\ij_x}  \ad_{\vec{j}} \a_{\vec{i}} ~ \hat{\tau}^z_{\ij_x} + \hc \r  \\
-  \sum_{\ij_y} \bigg[ t^a_y  ~\hat{\lambda}^y \l \ad_{\vec{j}} \a_{\vec{i}} \hat{\tau}^z_{\ij_y} + \hc \r + t_y^f ~  \hat{\Lambda}^y_{\ij_y} \hat{\tau}^x_{\ij_y} \bigg],
\label{eqHeffLadderGI}
\end{multline}
by fixing the additional link variables and working in the sector with $\hat{\tau}^z_{\ij_x} \equiv 1$. Note that the resulting physics is unaffected because the additional link variables are all conserved quantities, $[\tilde{\mathcal{H}}_{\rm 2leg}  , \hat{\tau}^z_{\ij_x}] =0$. To obtain the generalized gauge-invariant expressions for $\hat{\lambda}^\mu_{\ij_x}$ from Eq.~\eqref{eqDeflbdax}, $\hat{B}_p$ is replaced by the 2D plaquette term,
\begin{equation}
\hat{B}_p = \prod_{\ij \in \partial p} \hat{\tau}^z_{\ij},
\label{eqDefBpGI}
\end{equation}
which includes link variables $\hat{\tau}^z_{\ij}$ on the legs of the ladder. Similarly, a generalization of Eq.~\eqref{eqDefHLGT} is obtained,
\begin{multline}
\tilde{\mathcal{H}}_{\rm 2leg}^{\rm simp} = -  \sum_{\ij_x}  \tilde{t}^a_x \l \ad_{\vec{j}} \a_{\vec{i}}  \hat{\tau}_{\ij_x}^z + \hc \r \\
- \sum_{\ij_y} \left[ \tilde{t}^a_y \l \ad_{\vec{j}} \a_{\vec{i}}  \hat{\tau}_{\ij_y}^z + \hc \r  + \tilde{t}^f_y ~ \hat{\tau}_{\ij_y}^x \right].
\label{eqDefHLGTGI}
\end{multline}

\emph{Symmetries.--}
The operators $\hat{Q}_{\vec{j}}$ and $\hat{B}_p$ are both invariant under unitary $\Zt$ gauge transformations in 2D, defined by
\begin{equation}
\hat{G}_{\vec{j}} = \hat{Q}_{\vec{j}} \prod_{\vec{i}: \langle \vec{j}, \vec{i} \rangle} \hat{\tau}^x_{\langle \vec{j}, \vec{i} \rangle},
\label{eqDefGjLdr}
\end{equation}
where the product on the right includes all links $\langle \vec{j}, \vec{i} \rangle$ connected to site $\vec{j}$. By extending the Hilbert space and allowing arbitrary configurations $\hat{\tau}^z_{\ij_x} = \pm 1$ of the additional link variables on the legs, we promote Eq.~\eqref{eqHeffLadder} to a genuine $\Zt$ LGT characterized by local instead of global symmetries: we confirm below that $[\tilde{\mathcal{H}}_{\rm 2leg}, \hat{G}_{\vec{j}}] = 0$ and $[\tilde{\mathcal{H}}_{\rm 2leg}^{\rm simp} , \hat{G}_{\vec{j}}] = 0$ for all $\vec{j}$. 

\emph{$U(1)$ and $\Zt$ gauge structures.--}
The simplest cold-atom setups, with link variables $\tau^z_{\ij_y} = \pm 1$ on the rungs only, implement directly the sector $\hat{\tau}^z_{\ij_x} \equiv 1$ of the extended model \eqref{eqHeffLadderGI}. If, on the other hand, we want to realize Eq.~\eqref{eqHeffLadderGI} in a different sector of the Hilbert space -- e.g. where the $\Zt$ Gauss law $\hat{G}_{\vec{j}} \ket{\psi} = 1$ for all $\vec{j}$ is satisfied -- we need to involve basis states with $\hat{\tau}^z_{\ij_x} = \pm 1$. 

Further below we make use of the local $U(1)$ gauge invariance of the experimentally more easily realizable model \eqref{eqHeffLadder} and show that all configurations $\hat{\tau}^z_{\ij_x} = \pm 1$ are equivalent up to re-labelings $\a_{\vec{j}} \to \pm \a_{\vec{j}}$ and $\hat{\tau}^z_{\ij_y} \to \pm \hat{\tau}^z_{\ij_y}$ on the rungs. We show explicitly that all relevant observables of the model \eqref{eqHeffLadderGI} with the enlarged Hilbert space can be accessed experimentally by realizing only the sector with $\hat{\tau}^z_{\ij_x} \equiv 1$ on the legs. In this sense, the physical properties of the fully gauge-invariant model Eq.~\eqref{eqHeffLadderGI} can be accessed by implementing the simpler model in Eq.~\eqref{eqHeffLadder}.

\emph{$\Zt$ gauge invariant order parameters.--}
Next we formulate the order parameters of the original model \eqref{eqDefHLGT} in a fully gauge-invariant way. As shown explicitly below, a measurement of $\langle \hat{\tau}_{\ij_y}^z \rangle$ in the sector where $\hat{\tau}^z_{\ij_x}=1$, corresponds to a measurement of a $\Zt$ Wilson line $\langle \hat{W}_{\vec{j}} \rangle$ in the gauge invariant sector where the $\Zt$ Gauss law is satisfied, i.e. $\hat{G}_{\vec{j}} \ket{\psi} = 1$ for all $\vec{j}$. The Wilson line,
\begin{equation}
\hat{W}_{\vec{j}}= \prod_{\langle \vec{k}, \vec{i} \rangle \in \mathcal{C}_{j_x}} \hat{\tau}^z_{\langle \vec{k}, \vec{i} \rangle},
\label{eqDefZtWilsonLine}
\end{equation}
is defined as a product over all links $\langle \vec{k}, \vec{i} \rangle$ along a contour $\mathcal{C}_{j_x}$; specifically, we consider contours which start and end at the two sites on one of the edges of the ladder and which are oriented along $x$ except at a single bond with $x$-coordinate $j_x$ where $\mathcal{C}_{j_x}$ points along $y$-direction. Indeed, after setting $ \hat{\tau}_{\langle \vec{k}, \vec{i} \rangle_x}^z=1$ on all links $\langle \vec{k}, \vec{i} \rangle_x$ along $x$ in Eq.~\eqref{eqDefZtWilsonLine}, $\hat{W}_{\vec{j}} =  \hat{\tau}_{\ij_y}^z$ reduces to the $\Zt$ gauge field on link $\ij_y$.

Hence, the non-trivial phase in the gauge field sector is associated with a non-local string order parameter -- the $\Zt$ Wilson line -- when the $\Zt$ Gauss law is satisfied. Fully closed Wilson loops in the gauge-invariant model \eqref{eqDefHLGTGI} can be treated in a similar manner. They can be experimentally accessed in state-of-the-art cold atom experiments in the sector with $\hat{\tau}^z_{\ij_x}=1$ on the legs, by measuring correlation functions $W(d) = \langle \hat{\tau}^z_{\ij_y} \hat{\tau}^z_{\langle \vec{i}+d \vec{e}_x , \vec{j} +d \vec{e}_x \rangle_y} \rangle$. These observables are discussed in the main text.

\textbf{$\Zt$ gauge invariance.}
Now we provide a proof that Eq.~\eqref{eqHeffLadderGI} is invariant under the $\Zt$ gauge transformations $\hat{G}_{\vec{j}}$ introduced in Eq.~\eqref{eqDefGjLdr}. This property directly carries over to Eq.~\eqref{eqDefHLGTGI}, which is a special case of the more general Hamiltonian \eqref{eqHeffLadderGI}. First we note that the plaquette operators $\hat{B}_p$, see Eq.~\eqref{eqDefBpGI}, are $\Zt$ gauge invariant, 
\begin{equation}
[\hat{B}_p,\hat{G}_{\vec{j}}] = 0
\label{eqBpGaugeInv}
\end{equation}
for all $p$ and $\vec{j}$. Every pair of operators $\hat{B}_p$ and $\hat{G}_{\vec{j}}$ involves either zero or two common links $\ell_{1,2}$. In the first case Eq.~\eqref{eqBpGaugeInv} is trivially true. In the second case one obtains products like $\hat{\tau}^z_{\ell_1} \hat{\tau}^z_{\ell_2} \hat{\tau}^x_{\ell_1} \hat{\tau}^x_{\ell_2} =\hat{\tau}^x_{\ell_1} \hat{\tau}^x_{\ell_2}  \hat{\tau}^z_{\ell_1} \hat{\tau}^z_{\ell_2} $, because Pauli matrices anti-commute. The $\Zt$ gauge invariant plaquette terms correspond to the $\Zt$ magnetic field. 

Using Eq.~\eqref{eqBpGaugeInv} it is easy to see that $[\hat{\lambda}_{\ij}^\mu , \hat{G}_{\vec{k}}] = 0$ and $[\hat{\Lambda}^y_{\ij} , \hat{G}_{\vec{k}}] = 0$ for all $\ij$ and $\vec{k}$, because $\hat{\lambda}^\mu_{\ij}$ and $\hat{\Lambda}^y_{\ij}$ depend only on the $\Zt$ gauge invariant magnetic field, $\hat{B}_p$, and charge, $\hat{Q}_{\vec{j}}$. Obviously $[\hat{\tau}^x_{\ij} , \hat{G}_{\vec{k}}] = 0$, because $ \hat{G}_{\vec{k}}$ is a function of $\hat{\tau}^x_{\langle \vec{k}, \vec{l} \rangle}$ and $\hat{Q}_{\vec{k}}$ only. Finally, the tunneling terms are also $\Zt$ gauge invariant, $[\ad_{\vec{i}} \a_{\vec{j}} \hat{\tau}^z_{\ij}, \hat{G}_{\vec{k}} ] = 0$, because the $\Zt$ charges on sites $\vec{i}$ and $\vec{j}$ change sign and $\hat{\tau}^z_{\ij}$ anti-commutes with $\tau^x_{\ij}$ appearing in $\hat{G}_{\vec{i}}$ and $\hat{G}_{\vec{j}}$. Combining these results, it follows that Eq.~\eqref{eqHeffLadderGI} is $\Zt$ gauge invariant. 

\textbf{Auxiliary link variables.}
In the main text we have suggested to realize the model Eq.~\eqref{eqHeffLadder} first and introduced additional auxiliary link variables $\hat{\tau}^z_{\ij_x}$ on the legs of the ladder by hand; in the physical model they are fixed to $\hat{\tau}^z_{\ij_x} \equiv 1$. Because the effective Hamiltonian commutes with $\hat{\tau}^z_{\ij_x}$, the additional link variables are conserved quantities and remain fixed during time evolution. Now we demonstrate that the model \eqref{eqHeffLadderGI} is experimentally accessible for arbitrary configurations $\hat{\tau}^z_{\ij_x} = \pm 1$ on the legs, i.e., not only for $\hat{\tau}^z_{\ij_x} \equiv 1$ which can be directly implemented experimentally. 

Consider the effective Hamiltonian $\tilde{\mathcal{H}}_{\rm 2leg}[\tau^z_{\ij_x}]$ from Eq.~\eqref{eqHeffLadderGI} for a given configuration of the link variables on the legs, $\tau^z_{\ij_x} = \pm 1$. We will now explicitly construct a unitary transformation $\hat{U}$ for which 
\begin{equation}
\hat{U}^\dagger \tilde{\mathcal{H}}_{\rm 2leg}[\tau^z_{\ij_x}] \hat{U}  = \tilde{\mathcal{H}}_{\rm 2leg}[\sigma^z_{\ij_x}] , \quad \sigma^z_{\ij_x} \equiv 1.
\label{eqBasisChangeArbtryGauge}
\end{equation}
As discussed previously, the Hamiltonian on the right hand side can be directly accessed experimentally. Hence, by applying the unitary basis transformation $\hat{U}$ to the results of a measurement in the experimental basis allows to access observables for the Hamiltonian $ \tilde{\mathcal{H}}_{\rm 2leg}[\tau^z_{\ij_x}]$.

To construct $\hat{U}$, we first define string variables
\begin{equation}
\Sigma_\nu(j_x) = \prod_{i_x < j_x} \tau^z_{\langle ( i_x+1, \nu) , ( i_x , \nu) \rangle_x} = \pm 1,
\end{equation}
where $\nu =0,1$ denotes the $y$ coordinate of the leg on which the string is defined. Next we perform $U(1)$ gauge transformations on the $a$-particles,
\begin{equation}
\hat{U}_\nu^{U(1)} = \prod_{j_x}  ( \Sigma_\nu(j_x) )^{\hat{n}^a_{(j_x,\nu)}},
\end{equation}
such that 
\begin{equation}
(\hat{U}_\nu^{U(1)} )^\dagger ~ \hat{a}_{(j_x,\nu)} ~ \hat{U}_\nu^{U(1)} = \Sigma_\nu(j_x)~ \hat{a}_{(j_x,\nu)}.
\end{equation}
This is sufficient to bring the hopping terms of the matter field $\a$ on the legs to the desired form,
\begin{multline}
(\hat{U}_\nu^{U(1)} )^\dagger ~ \ad_{(j_x+1,\nu)} \a_{(j_x,\nu)} \tau^z_{\langle (j_x+1,\nu), (j_x,\nu) \rangle_x} ~ \hat{U}_\nu^{U(1)}\\
 =  \ad_{(j_x+1,\nu)} \a_{(j_x,\nu)},
\end{multline}
because $\Sigma_\nu(j_x+1) \Sigma_\nu(j_x) = \tau^z_{\langle (j_x+1,\nu), (j_x,\nu) \rangle_x}$ and making use of $(\tau^z_{\langle (j_x+1,\nu), (j_x,\nu) \rangle_x})^2=1$.

The $U(1)$ gauge transformations, defined as $\hat{U}^{U(1)} = \prod_{\nu=0,1} \hat{U}^{U(1)}_\nu$, also change the hoppings on the rungs,
\begin{multline}
(\hat{U}^{U(1)} )^\dagger ~ \ad_{(j_x,1)} \a_{(j_x,0)} \tau^z_{\langle (j_x,1), (j_x,0) \rangle_y} ~ \hat{U}^{U(1)}\\
 = \Sigma_1(j_x) \Sigma_0(j_x) ~ \ad_{(j_x,1)} \a_{(j_x,0)} \tau^z_{\langle (j_x,1), (j_x,0) \rangle_y}.
\end{multline}
To cancel the additional term $\Sigma_1(j_x) \Sigma_0(j_x)$, we perform another basis transformation, this time involving the link variables on the rungs,
\begin{equation}
\hat{V} = \prod_{j_x} \exp \left[ i \frac{\pi}{4} \bigl( 1 - \prod_\nu \Sigma_n(j_x) \bigr) ~ \hat{\tau}^x_{\langle (j_x,1), (j_x,0) \rangle_y} \right].
\end{equation}
This transformation leads to
\begin{equation}
\hat{V}^\dagger \hat{\tau}^z_{\langle (j_x,1), (j_x,0) \rangle_y} \hat{V} = \Sigma_1(j_x) \Sigma_0(j_x) ~ \hat{\tau}^z_{\langle (j_x,1), (j_x,0) \rangle_y},
\end{equation}
i.e. the link variable on the rung changes sign if $\Sigma_1(j_x) \Sigma_0(j_x) = -1$.

Combining these expressions, we obtain the desired result in Eq.~\eqref{eqBasisChangeArbtryGauge}: The unitary transformation is
\begin{equation}
\hat{U} = \hat{V} \prod_{\nu=0,1} \hat{U}^{U(1)}_\nu,
\label{eqDefU}
\end{equation}
and it consists of a combination of $U(1)$ gauge transformations of the $a$-particles and sign-changes of the link variables on the rungs,
\begin{equation}
\hat{U}^\dagger \hat{\tau}^z_{\ij_y} \hat{U} = \pm \hat{\tau}^z_{\ij_y},
\end{equation}
depending explicitly on the configuration of the auxiliary link variables $\tau^z_{\ij_x}$ on the legs.

\textbf{Accessing physical observables.}
Now we demonstrate how the unitary transformation \eqref{eqDefU} allows to measure the most important $\Zt$ gauge invariant quantities in the subspace of states $\ket{\psi}$ obeying the $\Zt$ Gauss law,
\begin{equation}
\hat{G}_{\vec{j}} \ket{\psi} = 1, \quad \forall \vec{j}.
\label{eqMethGaussLaw}
\end{equation}
We show that the expectation values of certain observables specified below, $\langle \hat{O} \rangle$, can be obtained, when only the subspace with $\hat{\tau}^z_{\ij_x} = 1$ is experimentally accessible.

For concreteness, we consider a ground state $\ket{\Psi_0}$ of Eq.~\eqref{eqHeffLadderGI} which satisfies $\hat{G}_{\vec{j}} \ket{\Psi_0} = 1$ for all sites $\vec{j}$. Because $\hat{\tau}^z_{\ij_x}$ and $\hat{G}_{\vec{k}}$ do not commute in general, the ground state $\ket{\Psi_0}$ becomes a superposition of all possible configurations $\tau^z_{\ij_x}$ of the auxiliary link variables on the legs,
\begin{equation}
\ket{\Psi_0} = \sum_{\tau^z_{\ij_x}} \Psi_0[\tau^z_{\ij_x}] ~ \ket{\tau^z_{\ij_x}} \otimes \ket{\Phi[\tau^z_{\ij_x}]}.
\end{equation}
Here the wavefunction $\ket{\Phi}$ involves only the matter field $\a$ and the link variables on the rungs. Proper normalization requires $\sum_{\tau^z_{\ij_x}} |\Psi_0[\tau^z_{\ij_x}] |^2 = 1$ and $\bra{~ \Phi[\tau^z_{\ij_x}]~} ~ \Phi[\tau^z_{\ij_x}] ~ \rangle = 1$.

Next we make use of the unitary transformations $\hat{U}$ in Eq.~\eqref{eqDefU} to relate all wavefunctions $\ket{\Phi[\tau^z_{\ij_x}]}$, which still explicitly depend on the configuration $\tau^z_{\ij_x}$, to the ground state $\ket{\Phi_0} = \ket{\Phi[\sigma^z_{\ij_x}]}$, with $\sigma^z_{\ij_x} \equiv 1$ for all links $\ij_x$, corresponding to the experimentally relevant sector of the Hilbert space. After replacing the numbers $\tau^z_{\ij_x}$ in the definition of $\hat{U}$ above with operators $\hat{\tau}^z_{\ij_x}$, we obtain
\begin{equation}
\ket{\Psi_0} = \sum_{\tau^z_{\ij_x}} \Psi_0[\tau^z_{\ij_x}] ~ \hat{U} ~ \ket{\tau^z_{\ij_x}} \otimes \ket{\Phi_0}.
\label{eqPsi0FromTrivial}
\end{equation}

Now we will prove the following statement: \emph{Consider observables $\hat{O}$ in the extended Hilbert space (i.e. including link variables on the legs) for which
\begin{multline}
\bra{\tau^z_{\ij_x}} \bra{\Phi} ~ \hat{U}^\dagger \hat{O} \hat{U} ~  \ket{\Phi} \ket{\overline{\tau}^z_{\ij_x}} = \\
= \delta(\tau^z_{\ij_x}, \overline{\tau}^z_{\ij_x})  ~ \bra{\sigma^z_{\ij_x}} \bra{\Phi} ~ \hat{O} ~ \ket{\Phi} \ket{\sigma^z_{\ij_x}},
\label{eqCondOmeasurable}
\end{multline}
with $\sigma^z_{\ij_x} \equiv 1$ the trivial configuration. Then 
\begin{equation}
\bra{\Psi_0} ~ \hat{O} ~ \ket{\Psi_0} =  \bra{\sigma^z_{\ij_x}} \bra{\Phi_0} ~ \hat{O} ~ \ket{\Phi_0} \ket{\sigma^z_{\ij_x}},
\label{eqOmeasurable}
\end{equation}
i.e. the observable $\hat{O}$ can be measured in the experimentally accessible part of the Hilbert space with $\hat{\tau}^z_{\ij_x} \equiv 1$.
}

For the proof we express $\bra{\Psi_0} ~ \hat{O} ~ \ket{\Psi_0}$ using Eq.~\eqref{eqPsi0FromTrivial}. Because of the Dirac delta function in Eq.~\eqref{eqCondOmeasurable}, we obtain only one sum over all configurations $\tau^z_{\ij_x}$ of the auxiliary link variables on the legs. Using the normalization condition for the amplitudes $|\Psi_0[\tau^z_{\ij_x}]|$ introduced above, Eq.~\eqref{eqOmeasurable} follows immediately.

Now we apply the above theorem and show that the following important observables can be measured in the experimentally accessible part of the Hilbert space with $\hat{\tau}^z_{\ij_x} \equiv 1$ on the legs of the ladder: 
\begin{itemize}
\item[(i)] The $\Zt$ electric field on the rungs, $\hat{O} = \hat{\tau}^x_{\ij_y}$.
\item[(ii)] The number density $\hat{O} = \hat{n}^a_{\vec{j}}$ of the matter field.
\item[(iii)] The $\Zt$ magnetic field (plaquette terms), $\hat{O} = \hat{B}_p$.
\item[(iv)] $\Zt$ gauge invariant rung tunneling, $\hat{O} = \hat{\tau}^z_{\ij_y} \ad_{\vec{i}} \a_{\vec{j}}$. 
\item[(v)] $\Zt$ Wilson loops, $\hat{O} = \prod_{\langle \vec{k}, \vec{i} \rangle \in \mathcal{C}} \hat{\tau}^z_{\langle \vec{k}, \vec{i} \rangle}$ for contours $\mathcal{C}$ which are closed or start at the ends of the ladder.
\end{itemize}

In all cases, we proof that condition \eqref{eqCondOmeasurable} is satisfied for the observables. For cases (i) and (ii) this is trivial, because $[\hat{O}, \hat{U}] = 0$ and $\hat{O}$ has no effect on the link variables on the legs. The other cases require more care. 

For (iii) consider a plaquette with sites $\vec{i}$, $\vec{j} = \vec{i} + \vec{e}_y$, $\vec{k} = \vec{i} + \vec{e}_x$ and $\vec{l} = \vec{j} + \vec{e}_x$. Next we note that, by construction of the unitary transformation $\hat{U}$,
\begin{flalign}
\hat{U}^\dagger \hat{\tau}^z_{\ij_y} \hat{\tau}^z_{\langle \vec{k}, \vec{l} \rangle_y} \hat{U} &= \hat{B}_{p(\langle \vec{i} , \vec{k} \rangle_x)}, \\
\hat{U}^\dagger \hat{\tau}^z_{\langle \vec{k}, \vec{i} \rangle_x} \hat{\tau}^z_{\langle \vec{l}, \vec{j} \rangle_x} \hat{U} &= \hat{\tau}^z_{\langle \vec{k}, \vec{i} \rangle_x} \hat{\tau}^z_{\langle \vec{l}, \vec{j} \rangle_x}.
\end{flalign}
Using $(\hat{\tau}^z_{\langle \vec{k}, \vec{i} \rangle_x} \hat{\tau}^z_{\langle \vec{l}, \vec{j} \rangle_x})^2=1$ we thus obtain
\begin{multline}
\bra{\tau^z_{\ij_x}} \bra{\Phi} ~ \hat{U}^\dagger \hat{B}_{p(\langle \vec{i} , \vec{k} \rangle_x)} \hat{U} ~  \ket{\Phi} \ket{\overline{\tau}^z_{\ij_x}}\\
= \bra{\tau^z_{\ij_x}} \bra{\Phi} ~\hat{\tau}^z_{\ij_y} \hat{\tau}^z_{\langle \vec{k}, \vec{l} \rangle_y}  ~  \ket{\Phi} \ket{\overline{\tau}^z_{\ij_x}}  \\
= \delta(\tau^z_{\ij_x}, \overline{\tau}^z_{\ij_x}) ~  \bra{\Phi} ~\hat{\tau}^z_{\ij_y} \hat{\tau}^z_{\langle \vec{k}, \vec{l} \rangle_y}  ~  \ket{\Phi} \\
= \delta(\tau^z_{\ij_x}, \overline{\tau}^z_{\ij_x}) ~  \bra{\sigma^z_{\ij_x}} \bra{\Phi} ~ \hat{B}_{p(\langle \vec{i} , \vec{k} \rangle_x)} ~ \ket{\Phi} \ket{\sigma^z_{\ij_x}}.
\end{multline} 

\begin{figure}[t!]
\centering
\epsfig{file=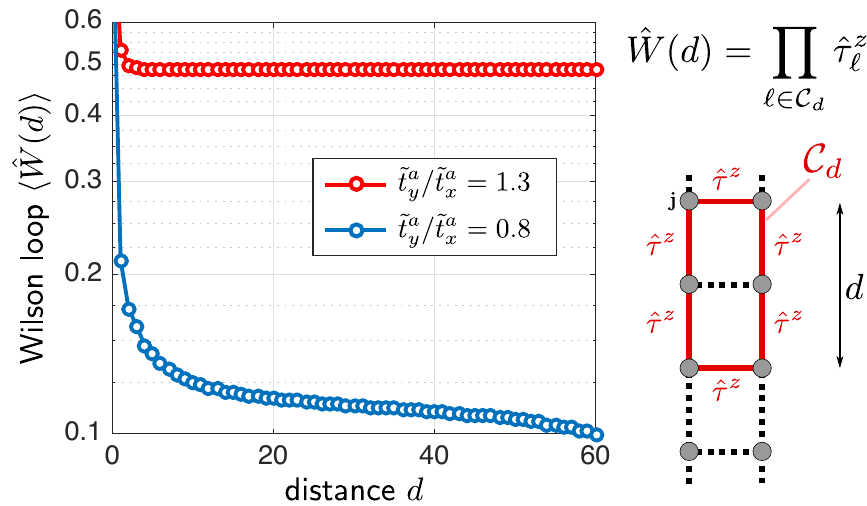, width=0.5\textwidth}
\caption{\textbf{Wilson loop scaling.} Using DMRG we calculate the ground state expectation value of the Wilson loop operator $\hat{W}(d)$, see sketch, for different sizes $d$ of the loop. The system has $L_x=96$ rungs and is in a rung-Mott phase, corresponding to the commensurate filling $N_a=L_x$ of the matter field $\a$; the ratio $\tilde{t}^f_y / \tilde{t}^a_x = 0.54$ is fixed, and we evaluated Wilson loops in the center of the system in order to avoid edge effects.}
\label{figWilsonLoopScaling}
\end{figure}

To show (iv), we note that 
\begin{flalign}
\hat{U}^\dagger \hat{\tau}^z_{\ij_y} \hat{U} &= \hat{\Sigma}_0(j_x) \hat{\Sigma}_1(j_x) ~ \hat{\tau}^z_{\ij_y},\\
\hat{U}^\dagger ~ \ad_{\vec{i}} \a_{\vec{j}}  ~ \hat{U} &= \hat{\Sigma}_0(j_x) \hat{\Sigma}_1(j_x) ~ \ad_{\vec{i}} \a_{\vec{j}},
\end{flalign}
from which the rest follows easily.

The proof of (v) for closed contours $\mathcal{C}$ is similar to the proof of (iii). We consider the contour $\mathcal{C}_{j_x}$ starting at the end of the ladder in more detail; See Fig.~\ref{figPhaseTrans} (C) in the main text for an illustration. We start by noting that the relevant Wilson line $\hat{W}_{\vec{j}} =  \hat{\Sigma}_0(j_x) \hat{\tau}^z_{\ij_y} \hat{\Sigma}_1(j_x)$. Hence the effect of the unitary transformation $\hat{U}$ is 
\begin{equation}
\hat{U}^\dagger ~\hat{W}_{\vec{j}}  ~ \hat{U} = \l \hat{\Sigma}_0(j_x) \r^2 \hat{\tau}^z_{\ij_y} \l \hat{\Sigma}_1(j_x) \r^2 = \hat{\tau}^z_{\ij_y},
\end{equation}
because $(\hat{\Sigma}_i(j_x))^2=1$ for $i=1,2$. From here the result follows easily.

\textbf{Wilson loops.} 
In Fig.~\ref{figWilsonLoopScaling} we present numerical results for the Wilson loops. We applied the theorem from above and calculated expectation values $W(d) = \langle \hat{\tau}^z_{\ij} \hat{\tau}^z_{\langle \vec{i}+d \vec{e}_x, \vec{j}+d \vec{e}_x \rangle} \rangle$ in the sector with $\hat{\tau}^z_{\ell_x} = 1$ on links $\ell_x$ along $x$-direction. 

In the phase with a Wilson-line order parameter at $\tilde{t}^a_y / \tilde{t}^a_x = 1.3$ (top red curve in Fig.~\ref{figWilsonLoopScaling}), where the $\Zt$ magnetic field dominates, the Wilson loop $W(d)$ quickly converges to a finite value for distances $d \geq 5$. This is indicative of a deconfined phase of the $\Zt$ gauge field. In the disordered phase at $\tilde{t}^a_y / \tilde{t}^a_x = 0.8$ (blue in Fig.~\ref{figWilsonLoopScaling}), where the $\Zt$ electric field dominates, the Wilson loop $W(d)$ slowly decays to zero at large distances. This behavior is reminiscent of a confined phase of the $\Zt$ gauge field. From the results in Fig.~\ref{figWilsonLoopScaling}, it is difficult to determine conclusively if the decay of $W(d)$ at large distances follows a power-law, or has a weak exponential dependence.

 ~\\
\textbf{V. Phase transitions of gauge and matter fields}\\
In this section, we provide additional details about the observed phase transitions in the gauge and charge sectors of the two-leg ladder Hamiltonian in Eq.~\eqref{eqDefHLGT}. While these transitions appear independently of each other in certain parameter regimes, we identify limits in which their interplay becomes important and they merge into a single transition point. This behavior is reminiscent of higher-dimensional realizations of the $\Zt$ LGT \cite{Fradkin1979,Lammert1995,Senthil2000}.

\begin{figure}[b!]
\centering
\epsfig{file=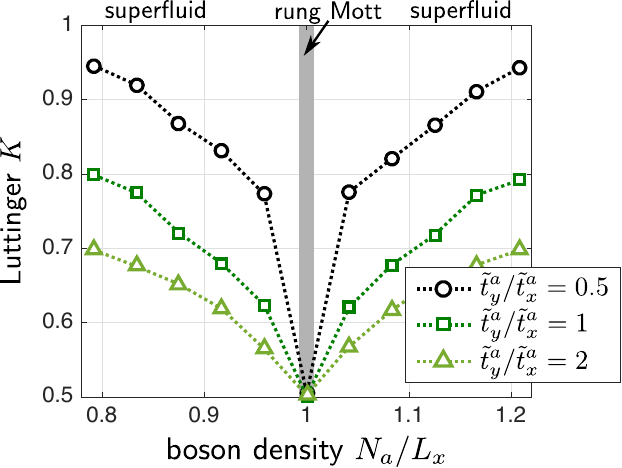, width=0.42\textwidth} $\qquad$
\caption{\textbf{The Luttinger-$K$ parameter,} as extracted from two-point correlators $g_0^{(1)}(i_x,j_x)$. In the immediate vicinity of the Mott transition, at the commensurate filling $N_a=L_x$, the Luttinger constant approaches $K=1/2$. Exactly at $N_a=L_x$ we extract the exponent from length scales smaller than the finite correlation length $\xi$ introduced by the Mott gap. We use DMRG simulations in system with $L_x=96$ rungs.}
\label{figLuttingerK}
\end{figure}
\begin{figure*}[t!]
\centering
\epsfig{file=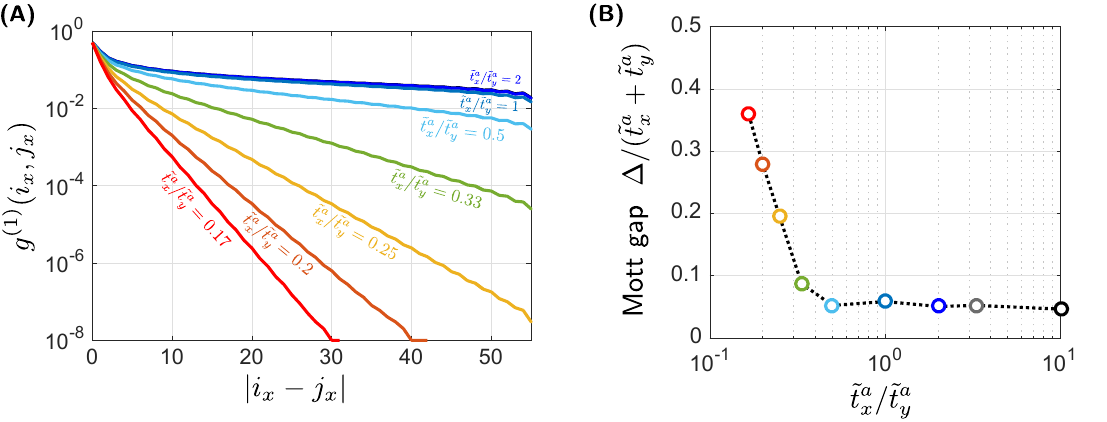, width=0.75\textwidth}
\caption{\textbf{Rung-Mott state at commensurate filling.} (A) The single-particle correlation function $g^{(1)}(i_x,j_x)$ is shown as a function of the distance $|i_x - j_x|$ at the commensurate filling $N_a = L_x$ and for different ratios of $\tilde{t}_x^a / \tilde{t}_y^a$; we fixed $\tilde{t}^f_y / \tilde{t}^a_y = 1$. For $\tilde{t}_x^a / \tilde{t}_y^a \lesssim 1/3$, clear exponential decay is found, as expected in an incompressible Mott state. (B) This picture is supported by the sizable Mott gap $\Delta$ in this regime. For larger values $\tilde{t}_x^a / \tilde{t}_y^a > 1/3$, the observed Mott gap is consistent with being caused entirely by finite-size effects. (A) We use DMRG simulations in a two-leg systems with $L_x=96$. (B) The TDL of the Mott gap is obtained by performing finite size extrapolation considering $L_x=N_a=24,36,48,60,72,84,96$.}
\label{figCommensurateMott}
\end{figure*}

\textbf{Superfluid-to-Mott transition.} 
In addition to the parity operator Eq.~\eqref{eqDefOP}, which we used in the main text to characterize the superfluid (SF) to Mott transition, we study the behavior of correlation functions at long distances. In the gapless SF regime, these are expected to have a power-law decay, with an exponent determined by the Luttinger constant $K$. We consider the single particle correlation function $g_0^{(1)}(i_x,j_x) = \langle \ad_{i_x \vec{e}_x} \a_{j_x \vec{e}_x} \rangle$. As discussed for example in Ref.~\cite{Crepin2011}, it is related to the Luttinger constant by $g_0^{(1)}(i_x,j_x) \sim |i_x-j_x|^{-\frac{1}{2K}}$. To obtain a fully $\Zt$ gauge-invariant observable, we introduce a string of $\hat{\tau}^z_{\ij_x}$ operators on the considered leg of the ladder. This allows to extract $K$ from 
\begin{multline}
g^{(1)}(i_x,j_x) = \langle \ad_{i_x \vec{e}_x} \biggl( \prod_{l_x=j_x}^{i_x-1} \hat{\tau}^z_{\langle (l_x+1) \vec{e}_x, l_x \vec{e}_x \rangle} \biggr) \a_{j_x \vec{e}_x} \rangle \\
\sim |i_x-j_x|^{-\frac{1}{2K}}
\label{eqTwoPtFnct}
\end{multline}
in the fully $\Zt$ gauge invariant formulation Eq.~\eqref{eqDefHLGTGI}.

Numerically, we find that the gauge-invariant two-point correlator $g_0^{(1)}(i_x,j_x)$ exhibits the expected power-law decay at large distances when $N_a \neq L_x$. The extracted Luttinger constants $K$ are shown in Fig.~\ref{figLuttingerK} for three different ratios $\tilde{t}^a_y / \tilde{t}^a_x$. At commensurate filling, $N_a=L_x$, all curves collapse at $K=1/2$. In this regime, and for sufficiently large values of $\tilde{t}^a_y / \tilde{t}^a_x$ (small values of $\tilde{t}^a_x / \tilde{t}^a_y$), we find that the correlation functions $g_0^{(1)}(i_x-j_x)$ decay exponentially at large distances, see Fig.~\ref{figCommensurateMott} (A). In these cases the Luttinger constant $K$ is determined by a fit to the correlation function at short distances, where the exponential decay has no effect yet. 

The observed exponential decay of the single-particle correlation function, as expected for a gapped Mott phase, is consistent with our results for the parity operator. The conclusion that the system is in an incompressible Mott state for sufficiently large values of $\tilde{t}^a_y / \tilde{t}^a_x$ is further supported by the value of the Luttinger constant, which approaches $K=1/2$ in the direct vicinity of the commensurate point. Indeed, this is the value of $K$ expected at the critical point where the Mott transition takes place \cite{Giamarchi2003}. These findings are also in very good agreement with the results obtained for a single component Bose system on a two-leg ladder, both in presence \cite{Piraud2015,DiDio2015,Romen2018} and in the absence \cite{Crepin2011} of synthetic magnetic flux. 

\begin{figure}[t!]
\centering
\epsfig{file=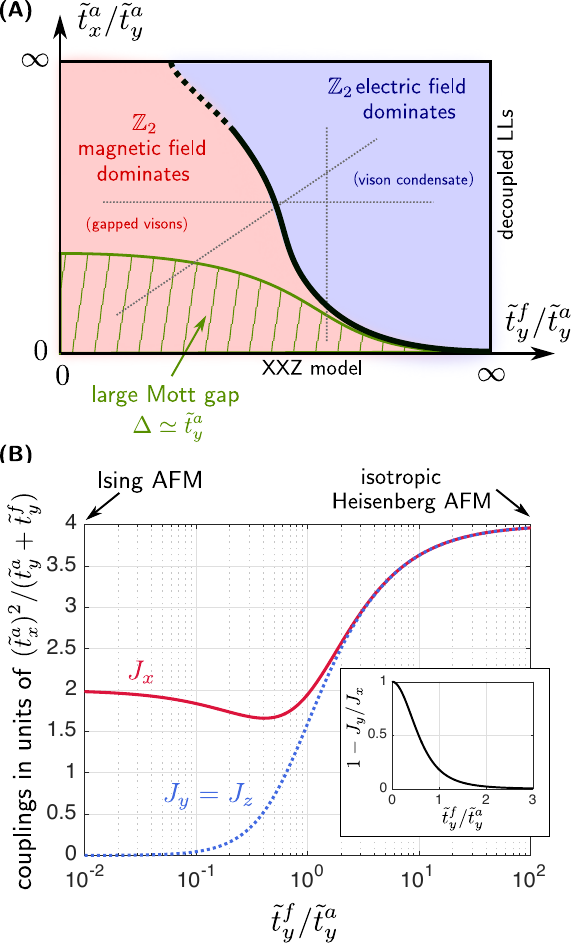, width=0.41\textwidth}
\caption{\textbf{Conjectured phase diagram of the $\Zt$ LGT on a two-leg ladder.} (A) We performed numerical DMRG simulations along the cuts indicated by gray lines. For small values of $\tilde{t}^f_y / \tilde{t}^a_y$ the system breaks the global $\Zt$ symmetry and is in an ordered phase dominated by the $\Zt$ magnetic field. Here we always expect a finite Mott gap, in analogy with models of hard-core bosons in a constant field \cite{Crepin2011,Piraud2015}. For larger values of $\tilde{t}^f_y / \tilde{t}^a_y$, the gauge field remains in a disordered phase and it is unclear whether the system remains in a Mott phase. Deep in the Mott regime (green shaded) the system can be mapped onto an effective spin-$1/2$ chain (B). It is described by an ${\rm XXZ}$ model with coupling constants $J_x$ and $J_y = J_z$ which are shown here as a function of $\tilde{t}^f_y / \tilde{t}^a_y$. In the inset we plot the anisotropy of the interactions, $(J_x - J_y) / J_x$, indicating that the ground state always breaks the global $\Zt$ symmetry deep in the Mott regime.}
\label{figGaugeMatterPhaseDiag}
\end{figure}

\textbf{The commensurate case $N_a = L_x$.} 
As explained in the main text, the transition in the gauge field sector corresponds to a spontaneous breaking of the global $\Zt$ symmetry, see Eq.~\eqref{eqZtSymm}. It is easy to diagnose from the order parameter $\langle \hat{\tau}^z_{\ij_y} \rangle \neq 0$. Now we study in more detail its interplay with the superfluid-to-Mott transition, which takes place at commensurate filling, $N_a = L_x$. 

In Fig.~\ref{figCommensurateMott} (A) we already found clearly exponentially decaying single-particle correlation functions when $\tilde{t}_x^a / \tilde{t}_y^a$ is small. For values $\tilde{t}_x^a / \tilde{t}_y^a > 1/3$ at $\tilde{t}^f_y = \tilde{t}^a_y$, the single-particle correlations are still clearly decaying, but the decay is also consistent with a power-law which would be expected in a gapless SF phase. A similar result is obtained in Fig.~\ref{figCommensurateMott} (B) where we study the Mott gap,
\begin{equation}
\Delta = E_0(N_a+1) + E_0(N_a-1) - 2 E_0(N_a).
\label{eqMottGapDef}
\end{equation}
While a sizable Mott gap is found for small values of $\tilde{t}_x^a / \tilde{t}_y^a$, the regime where $\tilde{t}_x^a / \tilde{t}_y^a$ is large is consistent with a compressible SF state or with an incompressible Mott state with a vanishingly small gap. The second scenario is indeed realized for hard-core bosons in a two-leg ladder without magnetic flux \cite{Crepin2011}. 

To gain more understanding of the phase diagram in the commensurate case, we performed different parameter scans as indicated in Fig.~\ref{figGaugeMatterPhaseDiag} (A). We found that for small values of $\tilde{t}^f_y / \tilde{t}^a_y$ the system is always in the ordered phase where the $\Zt$ magnetic field dominates. Around  $\tilde{t}^f_y = \tilde{t}^a_y$ we observe a transition to the disordered phase. In that regime we find that the Mott gap and the correlation functions are consistent with a gapless SF phase, or with a Mott phase with a small enough gap. The conjectured phase diagram is sketched in Fig.~\ref{figGaugeMatterPhaseDiag} (A). 

\textbf{Effective spin model in the Mott regime.} 
To shed more light on the transition in the gauge field sector and its interplay with the charge sector we now derive an effective low-energy spin model which is valid deep in the rung-Mott regime at the commensurate filling $N_a = L_x$. To this end we consider small values of $\tilde{t}^a_x \ll \Delta$, where $\Delta$ denotes the Mott gap, see Eq.~\eqref{eqMottGapDef}. 

Our starting point is the decoupled rung limit, $\tilde{t}^a_x = 0$. We consider the case when every rung is occupied by exactly one boson, $N_a=L_x$. Because of the $\Zt$ gauge invariance of the double-well system with sites $\vec{j}_{1,2}$, the single rung ground state is two-fold degenerate. The two ground states have energy $- \epsilon$, where
\begin{equation}
\epsilon = \sqrt{\l \tilde{t}^f_y \r^2 + \l \tilde{t}^a_y \r^2},
\end{equation}
and they can be written as
\begin{multline}
\ket{\! \uparrow} = \frac{1}{2\sqrt{\epsilon}} \biggl[ \frac{\tilde{t}^a_y}{\sqrt{ \epsilon + \tilde{t}^f_y }} ~ \biggl( \ket{{\rm L}, -} -  \ket{{\rm L}, +} \biggr)   \\
- \sqrt{\epsilon + \tilde{t}^f_y} ~  \biggl(  \ket{{\rm R}, +}  + \ket{{\rm R}, -} \biggr) \biggr], 
\label{eqDefUp}
\end{multline}
\begin{multline}
\ket{\! \downarrow}  = \frac{1}{2\sqrt{\epsilon}} \biggl[ \frac{\tilde{t}^a_y}{\sqrt{ \epsilon - \tilde{t}^f_y }} ~ \biggl( \ket{{\rm L}, -} +  \ket{{\rm L}, +} \biggr)   \\
+  \sqrt{\epsilon - \tilde{t}^f_y} ~  \biggl(  \ket{{\rm R}, +}  - \ket{{\rm R}, -} \biggr) \biggr]. 
\label{eqDefDown}
\end{multline}
Here $\ket{\mu,\tau^z}$, with $\mu = {\rm L,R}$ denotes the basis states with an $a$-particle on the left (L) site $\vec{j}_1$ [respectively, the right (R) site $\vec{j}_2$] and the gauge field in an eigenstate of $\hat{\tau}^z$ with eigenvalue $\tau^z = \pm 1$.

In the subspace spanned by the two spin states $\ket{\! \uparrow}$, $\ket{\! \downarrow}$ the $\hat{S}^z$ operator is identical to the generator $\hat{g}_{1}$ of $\Zt$ gauge transformations on the rung, defined in Eq.~\eqref{eqDefZtgaugeGroup2wll}: $\hat{g}_1 \ket{\! \uparrow} = \hat{S}^z \ket{\! \uparrow} = \ket{\! \uparrow}$ and $\hat{g}_1 \ket{\! \downarrow} = \hat{S}^z \ket{\! \downarrow} = - \ket{\! \downarrow}$. The two-fold degeneracy of the rung ground state can thus be understood as a consequence of the $\Zt$ gauge invariance of the decoupled rungs. 

The ground state energy of a state with two hard-core bosons on one rung is given by $- \tilde{t}^f_y$, because only the $f$-particles can tunnel between the two sites $\vec{j}_{1,2}$. Hence we obtain an asymptotic expression for the size of the Mott gap in the limit $\tilde{t}^a_x \to 0$,
\begin{equation}
\Delta =  \sqrt{\l \tilde{t}^f_y \r^2 + \l \tilde{t}^a_y \r^2} -  \tilde{t}^f_y.
\end{equation}
When $\tilde{t}^f_y < \tilde{t}^a_y$, we can thus approximate $\Delta \approx \tilde{t}^a_y$. When $\tilde{t}^f_y > \tilde{t}^a_y$ it follows $\Delta \approx 0.5 \l  \tilde{t}^a_y \r^2 / \tilde{t}^f_y$. These estimates determine where the rung-Mott state can be described as a product of singly-occupied rungs, each representing a localized magnetic moment with spin $S=1/2$ according to Eqs.~\eqref{eqDefUp}, \eqref{eqDefDown}. This corresponds to the green [striped] region illustrated schematically in Fig.~\ref{figGaugeMatterPhaseDiag} (A).  

To include the effects of tunneling $\tilde{t}^a_x$ along the legs of the ladder, we perform second order perturbation theory, effectively integrating out the virtual intermediate states with two bosons on one rung. This leads to an effective magnetic Hamiltonian, which can be formulated in terms of the spin-operators $\hat{S}^\alpha_j$ with $\alpha=x,y,z$ defined for the two spin states $\ket{\! \uparrow}$, $\ket{\! \downarrow}$ on rung $j=1...L_x$. To make our result more transparent, we perform an additional spin transformation and introduce
\begin{flalign}
\tilde{S}^x_j &= (-1)^j \hat{S}^x_j, \\
\tilde{S}^y_j &= (-1)^j \hat{S}^y_j, \\
\tilde{S}^z_j &= \hat{S}^z_j.
\end{flalign}

In terms of these operators, the effective Hamiltonian becomes an anti-ferromagnetic XXZ model, 
\begin{multline}
\H_{\rm eff} = - \l  \epsilon + \frac{J_x}{4} \r L_x + \\
+ \sum_{j=1}^{L_x} \biggl[ J_x \tilde{S}^x_{j+1} \tilde{S}^x_j + J_y \l \tilde{S}^y_{j+1} \tilde{S}^y_j + \tilde{S}^z_{j+1} \tilde{S}^z_j \r  \biggr].
\end{multline}
The coupling constants are given by
\begin{flalign}
J_x &= \l \tilde{t}^a_x \r^2 \left[ \eta + \l \tilde{t}^a_y \r^2 / \epsilon^3 \right] \geq 0, \\
J_y = J_z &= \l \tilde{t}^a_x \r^2 \left[ \eta - \l \tilde{t}^a_y \r^2 / \epsilon^3 \right] \geq 0,
\end{flalign}
where
\begin{equation}
\eta = \frac{\l 1 + \tilde{t}^f_y / \epsilon \r^2}{2 \l \epsilon - \tilde{t}^f_y \r } + \frac{\l 1 - \tilde{t}^f_y / \epsilon \r^2}{2 \l \epsilon + \tilde{t}^f_y \r }.
\end{equation}

The couplings $J_x$, $J_y = J_z$ are plotted as a function of $\tilde{t}^f_y / \tilde{t}^a_y$ in Fig.~\ref{figGaugeMatterPhaseDiag} (B). In the limiting cases $\tilde{t}^f_y / \tilde{t}^a_y= 0$, respectively $\tilde{t}^f_y  / \tilde{t}^a_y = \infty$, we obtain an Ising anti-ferromagnet (AFM), respectively an isotropic Heisenberg AFM. At intermediate values of $\tilde{t}^f_y / \tilde{t}^a_y$ the XXZ model has an Ising anisotropy with $J_x > J_{y} = J_z$, which becomes small when $\tilde{t}^f_y > \tilde{t}^a_y$, see inset in Fig.~\ref{figGaugeMatterPhaseDiag} (B).

The ground state of the obtained XXZ model with $J_x \geq J_y = J_z$ spontaneously breaks the discrete $\Zt$ symmetry, $\tilde{S}^x \to - \tilde{S}^x$ and $\tilde{S}^y \to - \tilde{S}^y$ but $\tilde{S}^z \to \tilde{S}^z$, unless $J_x=J_y = J_z$. In the symmetry-broken phase, the order parameter $(-1)^j \langle \tilde{S}^x_j \rangle = \langle \hat{S}^x_j \rangle$ corresponds to a non-vanishing expectation value of the $\Zt$ gauge field $\langle \hat{\tau}^z_{\ij_y} \rangle$ on the rungs. Therefore we conclude that the $\Zt$ LGT on the two-leg ladder is in the ordered phase, where the $\Zt$ magnetic field dominates, whenever the Mott gap is sizable. The transition to a disordered regime, where the $\Zt$ electric field dominates, is of Berezinskii-Kosterlitz-Thouless (BKT) type and takes place when $\tilde{t}^f_y / \tilde{t}^a_y \to \infty$. This is also where the Mott gap vanishes, and two decoupled SFs, described by Luttinger liquids (LLs), are obtained. These results, which we summarize in Fig.~\ref{figGaugeMatterPhaseDiag} (A), indicate an interesting interplay of the phase transitions in the gauge field and the charge sector when the filling fraction with bosons is commensurate, $N_a = L_x$.

~ \newpage
\textbf{\large References}
\vspace{-1.5cm}

~ \\
\textbf{Acknowledgements}

\textbf{General:} We would like to thank I. Bloch and M. Lohse for fruitful discussions. We also acknowledge discussions with P. Hauke, P. Zoller, V. Kasper, A. Bermudez, L. Santos, I. Carusotto and M. Hafezi.

\textbf{Funding:} The work in Brussels was supported by the FRS-FNRS (Belgium) and the ERC Starting Grant TopoCold. The work in Munich was supported by the Deutsche Forschungsgemeinschaft (FOR2414 Grant No. BL 574/17-1), the European Commission (UQUAM Grant No. 5319278), and the Nanosystems Initiative Munich (NIM) Grant No. EXC4. The work in Harvard was supported by the Gordon and Betty Moore foundation through the EPiQS program, Harvard-MIT CUA, NSF Grant No. DMR-1308435,
AFOSR-MURI Quantum Phases of Matter (grant FA9550-14-1-0035), and AFOSR-MURI: Photonic Quantum Matter (award FA95501610323). F.G. also acknowledges support from the Technical University of Munich - Institute for Advanced Study, funded by the German Excellence Initiative and the European Union FP7 under grant agreement 291763, from the DFG grant No. KN 1254/1-1, and DFG TRR80 (Project F8). 

\textbf{Author contributions} F.G. and N.G. devised the initial concepts. F.G. performed the main analytical calculations, with inputs from N.G. and E.D. All DMRG simulations were performed by L.B. The proposed experimental implementation was devised by C.S., M.A., N.G. and F.G. All authors contributed substantially to the analysis of the theoretical results. The manuscript was prepared by F.G., N.G., L.B. and C.S., with inputs from all other authors.

\textbf{Competing financial interests:} The authors declare no competing financial interests.

\textbf{Data availability:} The data that support the findings of this study are available from the corresponding author upon reasonable request.

\end{document}